# An International Survey of Front-End Receivers and Observing Performance of Telescopes for Radio Astronomy

P. Bolli[1], A. Orfei[2], A. Zanichelli[2], R. Prestage[3], S.J. Tingay[4], M. Beltrán[1], M. Burgay[5], C. Contavalle[2], M. Honma[6], A. Kraus[7], M. Lindqvist[8], J. Lopez Perez[9], P. Marongiu[5], T. Minamidani[10], S. Navarro[11], T. Pisanu[5], Z.-Q. Shen[12], B.W. Sohn[13], C. Stanghellini[2], T. Tzioumis[14], G. Zacchiroli[2]

[1] INAF-OAA, Istituto Nazionale di Astrofisica - Osservatorio Astrofisico di Arcetri (Firenze - Italy)
[2] INAF-IRA, Istituto Nazionale di Astrofisica - Istituto di Radioastronomia (Bologna - Italy)
[3] NRAO-GBT, National Radio Astronomy Observatory - Green Bank Telescope (Green Bank - USA)
[4] International Centre for Radio Astronomy Research, Curtin University (Bentley – Australia)
[5] INAF-OAC, Istituto Nazionale di Astrofisica - Osservatorio Astronomico di Cagliari (Cagliari - Italy)
[6] NAOJ, National Astronomical Observatory of Japan (Japan)
[7] MPIfR, Max Planck Institut für Radioastronomie (Bonn - Germany)
[8] OSO, Department of Space, Earth and Environment, Chalmers University of Technology, Onsala Space Observatory, 439 92 (Onsala, Sweden)
[9] IGN-CAY, Instituto Geográfico Nacional - Centro Astronómico de Yebes (Guadalajara - Spain)
[10] NAOJ-NRO, National Astronomical Observatory of Japan - Nobeyama Radio Observatory (Japan)
[11] IRAM, Instituto de Radioastronomía Milimétrica (Granada, España)
[12] CAS-ShAO, Chinese Academy of Sciences - Shanghai Astronomical Observatory (Shanghai - China)
[13] KASI, Korea Astronomy and Space Science Institute (Daejeon - Rep. of Korea)
[14] CSIRO-ATNF, Commonwealth Scientific and Industrial Research Organization - Australia Telescope National Facility (Epping - Australia)

**Abstract**

*This paper presents a survey of microwave front-end receivers installed at radio telescopes throughout the World. This unprecedented analysis was conducted as part of a review of front-end developments for Italian radio telescopes, initiated by the Italian National Institute for Astrophysics in 2016. Fifteen international radio telescopes have been selected to be representative of the instrumentation used for radio astronomical observations in the frequency domain from 300 MHz to 116 GHz. A comprehensive description of the existing receivers is presented and their characteristics are compared and discussed. The observing performances of the complete receiving chains are also presented. An overview of on-going developments illustrates and anticipates future trends in front-end projects to meet the most ambitious scientific research goals.*

## 1. Introduction

Front-end receivers represent one of the more sensitive and crucial components in the receiving scheme of a radio telescope. They are installed at the radio telescope foci and are designed to convert the electromagnetic waves from free-space to guided components, divide different orthogonal polarizations, and eventually amplify the sky radio frequency signal and down-convert it to lower frequency. After that, the back-end components, which represent another fundamental technological area, digitize the signal for further processing according to the specific astronomical requirements.

Front-end receivers have a lifetime that is usually shorter than the radio telescope itself, therefore keeping the receiver fleet aligned to the most advanced technological frontiers is a fundamental step to maintain the radio telescopes as scientifically competitive. This aspect is particularly relevant in the era of the largest radio astronomical facilities ever built by the international radio astronomical community, for instance the Square Kilometre Array (SKA) or the Atacama Large Millimeter Array (ALMA). The construction of new instrumentation at single-dish facilities can take advantage of the technological developments ongoing for these large international projects. Similarly, the availability of state-of-the-art receivers is crucial to allow scientists to perform cutting-edge research by exploiting the possible synergies among future world-class interferometers and single-dish radio telescopes.

Front-end developments have evolved significantly from the traditional mono-feed receivers operating in narrow frequency bands centered around specific emission spectral lines (which are still in operation to assure standard astronomical observations, for instance participation in Very Large Baseline Interferometry (VLBI) networks) to the most advanced and complex systems under development in the microwave laboratories of the radio astronomical observatories. After the generation of mono-feed receivers, the trend has been to add a second or even more feeds for improving the atmosphere calibration capabilities and for mapping large regions of sky. Two notable examples of this attempt to maximize the impact of large-format focal plane "radio cameras" are the APRICOT (All Purpose Radio Imaging Cameras on Telescopes) project funded by European Union within Cycle 2 of RadioNet FP7[1] and the dual-polarization 4x4 cryogenic focal plane array SEQUOIA built at the University of Massachusetts and currently in operation at the Large Millimeter Telescope.

Besides these developments, significant efforts have been made to improve the performance of each single microwave component making up the front-end chain, for instance reducing the receiver noise temperature and increasing the frequency bandwidth. Nowadays, modern astronomy pushes to open new frontiers in receiver technology, which can be mainly identified in two different areas: Ultra Wide Band (UWB) and Phased Array Feed (PAF) receivers. An example of the former is addressed by the BRoad bAND European VLBI Network (BRAND EVN) activity, under the umbrella of the RadioNet H2020 project, aimed at the development and construction of a prototype broad-band digital receiver in the frequency range 1.5 - 15.5 GHz (Tuccari et al. 2017). Many technological research projects to develop PAF receivers are underway, and first examples are installed at some observing facilities, such as APERTIF and ASKAP. The importance of PAF front-ends has increased in recent years due to their possible use for the SKA telescope.

Aimed at optimizing future receiver developments at the Italian National Institute for Astrophysics (INAF), in the period 2016 – 2017 an internal review was conducted of the radio astronomical front-end receivers for the INAF radio telescopes (Bolli et al. 2017, 2018). This review was part of a more general process within INAF aimed at harmonizing and coordinating efforts and resources in Italian radio astronomy. INAF operates three major, fully steerable radio astronomical instruments: the 64-m Sardinia Radio Telescope (SRT); and the 32-m Medicina and Noto Cassegrain radio telescopes. The target of the front-end review

---

included three main areas: *i)* the existing receiver systems and in particular those requiring maintenance/repair; *ii)* the receiver developments currently underway; and finally *iii)* a roadmap for future receiver developments.

As part of this review, twelve recognized top-class international radio astronomy facilities were contacted to survey the technical characteristics of their receivers in operation and under development. This gave the opportunity to put the technical characteristics of the Italian radio telescopes into this international context, to learn where INAF should seek greater levels of compatibility with other international facilities, and the directions future developments at INAF telescopes should take. The international radio telescopes included in the survey were selected based on the following criteria: *i)* facilities mainly used for single-dish observations and of similar class, in terms of size, to the Italian antennas; *ii)* radio telescopes operating in frequency ranges overlapping, at least partially, the coverage of the Italian radio telescopes; and *iii)* established contacts between INAF and the local staff. Additionally, the following couple of interferometers regularly used in combination with the Italian radio telescopes for interferometric observations were included as well: KVN (Korean VLBI Network); and VERA (VLBI Exploration of Radio Astrometry) formed by three and four radio telescopes, respectively. For these interferometric facilities, the properties of the individual antennas were taken into account. For radio telescopes exceeding the maximum frequency of the Italian radio telescopes (116 GHz), only the front-ends operating below this threshold were considered. Of course, many other telescopes may have been added; however, the variety of technical characteristics of the considered radio telescopes and the amount of data that we collected was sufficient to give a realistic picture of front-ends for radio astronomy worldwide.

More than 100 front-end receivers in operation at different radio telescopes have been surveyed, while approximately 20 future receivers have been considered. This leads to an impressive amount of data collected and analyzed during the survey, which are divided in three main categories: technical; scientific; and management. In particular, the key technical parameters identified as the most interesting for each receiver are represented by the frequency range, the feed-system category, the instantaneous bandwidth, the system noise temperature, the antenna gain, and the construction date.

The data collected during the international survey did not only provide useful information for the internal review. They were also a valuable set of coherent information able to provide a picture of the status and perspectives on front-end developments around the world. In particular, a detailed analysis of this set of data allowed us to draw a comparison among front-end receivers installed in various radio telescopes. Furthermore, an assessment of the observing performance of different telescopes requires consideration of other fundamental factors. To this end, our analysis includes antenna parameters and site characteristics. Moreover, the effect of the back-end components on telescope sensitivity has been taken into account in terms of processed instantaneous bandwidth. Finally, this survey also reports the ongoing front-end development at the various observatories. We hope that our investigation will prove useful to engineers in assessing the current and future state of the art in receiver design.  We also hope the survey will be useful to scientists, in quickly assessing in a single publication the facilities and receivers that could be used for their scientific programs.

The remainder of this paper is organized as follows. Section 2 lists the characteristics of the radio telescope included in the analysis, while section 3 gives some example science from each facility. Sections 4 and 5 illustrate the technical characteristics of the existing front-ends and the observing performance when installed at these radio telescopes respectively. Finally, section 6 describes future receiver developments, while our conclusions are summarized in section 7.

## 2.    Technical characteristics of the Radio Telescopes

The international radio telescopes selected in the review are located almost uniformly around the World: 8 in Europe including the 3 Italian radio telescopes; 4 in Asia; 2 in Australia; and 1 in the United States (see also Fig. 1). Table 1 lists some basic technical information, like nominal frequency range, altitude above sea level, and diameter of the main mirror. The abbreviations introduced in this table will be used throughout the article to identify the radio telescopes. Based on the antenna diameter, the radio telescopes have been divided in two main classes: large- (≥ 64 m) and medium-size (< 64 m). From Table 1, it can be seen that 3 mm observations are exploited even at some low altitude sites.

The radio telescopes considered in this survey have different optical configurations, ranging from single reflector to dual reflector and also beam wave guide type. The optical systems vary from Cassegrain to Gregorian or Nasmyth, in some cases with a shaped configuration of the mirrors. Active surfaces are offered by the more recent antennas and have been installed as an upgrade in some of the older antennas. Such a facility allows for the compensation of gravitational deformations of the telescope structure as well as non-systematic effects like wind and temperature gradients. Furthermore, all the facilities are equipped with frequency agility capabilities to place remotely, automatically, and quickly (typically within a few minutes) the desired receiver in the focal position. Changing receivers without manual intervention also improves the overall reliability of the system, because no disconnection of cables and removal of receivers are necessary.

| Radio telescope (abbreviation) | D (m) | Class | Nation | Altitude (m) | Frequency range (GHz) |
|---|---|---|---|---|---|
| Green Bank (GBT) | 100 | Large | USA | 807 | 0.3 - 116 |
| Effelsberg (EFF) | 100 | Large | Germany | 319 | 0.3 - 96 |
| Tianma (TMRT) | 65 | Large | China | 7 | 1.25 - 50 |
| Parkes (PARKES) | 64 | Large | Australia | 415 | 0.7 - 25 |
| Sardinia (SRT) | 64 | Large | Italy | 600 | 0.3 - 116 |
| Nobeyama (NOB) | 45 | Medium | Japan | 1349 | 20 - 116 |
| Yebes (YEBES) | 40 | Medium | Spain | 931 | 2 - 116 |
| Medicina (MED) | 32 | Medium | Italy | 25 | 1.35 - 26.5 |
| Noto (NOTO) | 32 | Medium | Italy | 78 | 1.4 - 86 |
| Pico Veleta (IRAM30) | 30 | Medium | Spain | 2850 | 70 - 370 |
| 25m Onsala (ONS25) | 25 | Medium | Sweden | 20 | 1.2 - 6.7 |
| Mopra (MOPRA) | 22 | Medium | Australia | 860 | 1.2 - 117 |
| Korean VLBI Network (KVN) | 21 | Medium | Korea | 120;260;320 | 21 - 142 |
| VLBI Exploration of Radio Astrometry (VERA) | 20 | Medium | Japan | 116;574;273;65 | 2.2 - 50 |
| 20m Onsala (ONS20) | 20 | Medium | Sweden | 20 | 2.2 - 116 |

Table 1 – Key-parameters of the radio telescopes included in the analysis. The order of the list is based on the diameter of the radio telescopes.

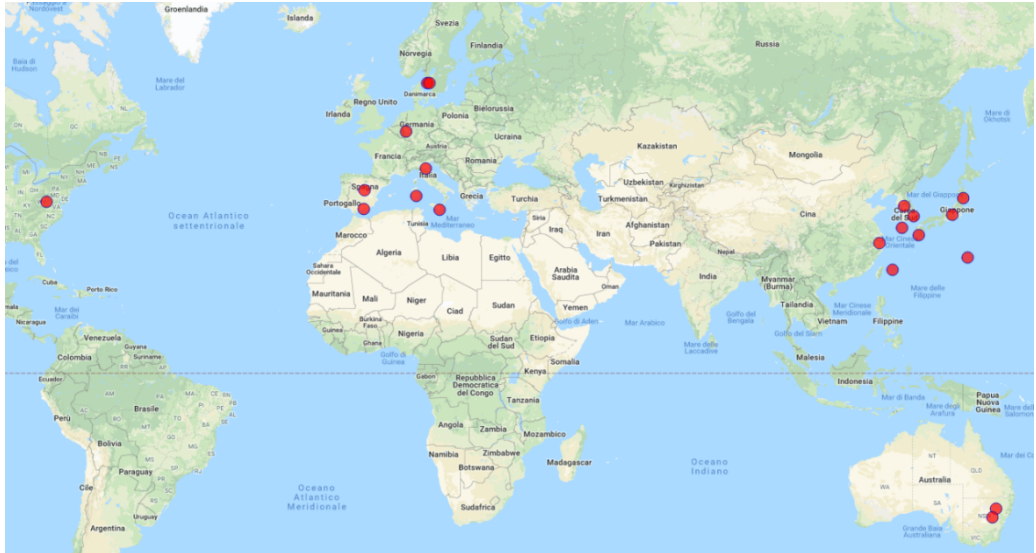

Figure 1 – Geographical distribution of the radio telescopes considered in the survey. The positions of all antennas of the KVN and VERA arrays are shown.

In the following sub-sections the main technical characteristics of each single radio telescope are described.

### 2.1. Green Bank Telescope  (USA)

The Robert C. Byrd Green Bank Telescope (Prestage et al. 2009) is a 100-m-diameter dual offset Gregorian reflector radio telescope operating in the frequency range ~300 MHz to 116 GHz. Its key technical features include: a large, unblocked aperture; a fully active surface of 2004 panels; a sophisticated system for real-time measurement and optimization of the wave front; and a highly advanced telescope control system that provides excellent pointing and tracking performance.

The GBT has an extremely flexible suite of instrumentation, including single- and dual-feed heterodyne receivers covering 290 MHz to 92 GHz. The telescope is currently equipped with four multi-pixel systems: FLAG, a 19 element, dual-polarization cryogenic PAF operating at 1.4 GHz that was in its commissioning phase at the time of this review (Roshi et al. 2018); KFPA, a seven pixel, dual-polarization feed horn array operating at 18 - 26 GHz (Morgan et al. 2008);  ARGUS, a 16-pixel single-polarization feed horn array operating in the 85-116 GHz range (Sieth et al. 2014); and MUSTANG-2, a 223-feed horn dual-polarization bolometer camera with 20 GHz bandwidth centered at 90 GHz (Stanchfield et al. 2016).

The main back-end is VEGAS (Versatile GBT Astronomical Spectrometer, Chennamangalam et al. 2014), an eight-bank spectrometer / pulsar / continuum back-end. Each bank can process up to 1.25 GHz of dual-polarization bandwidth; banks may be configured in serial (to obtain up to 10 GHz dual-polarization bandwidth), or in parallel (to observe up to 64 spectral windows within a 1.25 GHz passband). Additional back-ends support VLBI, radar, and other more specialized types of observation.

The GBT has a very flexible, and highly configurable observation control system. This supports all standard observing schemes, including various types of switching (position switching, frequency switching, sub-reflector nodding), pointed observations, grid maps, and on-the-fly mapping.  Advanced trajectory control, including the ability to perform lissajous figures and daisy-petal scans, optionally in an (Az, El) scanning frame about a J2000 tracking center, facilitate the use of the (in some cases spatially under-sampled) array receivers. GBT observations are typically dynamically scheduled 24 - 48 hours in advance (Balser et al. 2009) on the basis of internally generated, highly accurate three- and seven-day weather forecasts. The majority of observations are performed remotely, with the observer connecting to the GBT control system via a VNC session.

### 2.2. Effelsberg Radio Telescope (Germany)

The 100-m radio telescope (Hachenberg et al. 1973; Wielebinski et al. 2011) of the Max-Planck-Institut für Radioastronomie (MPIfR) is located in a valley near Bad Münstereifel-Effelsberg, about 40 km southwest of Bonn, to facilitate protection from Radio Frequency Interference (RFI). It can be used to observe radio emission from celestial objects in the frequency range 300 MHz to 96 GHz. The combination of the high surface accuracy of the reflector (~ 0.5 mm RMS) and the construction principle of "homologous distortion" (i.e., the reflector in any tilted position has a parabolic shape with a well-defined, but shifted, focal point) enables very sensitive observations even at the highest frequencies. Receivers are operated at the prime focus (mostly the long-wavelength systems) and also - due to its Gregorian optics with a 6.5 m elliptical secondary mirror - at the secondary focus of the antenna. With a nearly seamless (except for RFI polluted frequencies) frequency coverage from 1.4 GHz all the way up to 50 GHz, and the high frequency agility in the secondary focus (switching between receivers is done within 30 secs), the telescope provides high flexibility in the observation planning.

At the time of writing the receiver suite is undergoing a major upgrade with no interruption of the science operations. Systems with outdated performance are being replaced by new, state-of-the-art receivers. The new receivers are mostly broad-band systems with two feed horns, like the new K-band receiver which allows observations in the frequency range 18-26 GHz continuously. Among the new receivers there is also a PAF, which is a modified version of the one developed for the ASKAP telescope in Australia (Deng et al. 2017).

Receivers are combined with dedicated back-end systems for continuum observations (including polarimetry), spectroscopy, pulsar observations, and VLBI. Several FFT spectrometers with 65535 channels covering up to 2.5 GHz bandwidth are available, while for VLBI observations the telescope is equipped with Digital Base Band Converter (DBBC) and Roach Digital Back-End (RDBE) devices. Combining several spectrometers, it is possible to observe up to 8 GHz instantaneous bandwidth with ~1.8 million spectroscopic channels. A new spectro-polarimeter delivering the four Stokes parameters with 1024 channels is currently being commissioned at the telescope.

The EFF telescope control system allows tracking and scanning in various modes and different coordinate systems (horizontal, equatorial using various equinoxes, galactic and free-defined systems). Setting up observations is easily done by a Graphical User Interface (GUI), but it is also possible to use scripts and build up a queue of observations. Remote observations are possible through a VNC connection.

### 2.3. TianMa Radio Telescope  (China)

The TianMa Radio Telescope (Yan et al. 2015) is an omnidirectional rotatable radio telescope with a 65-m diameter main reflector. It is located at the Sheshan, Songjiang base of the Shanghai Astronomical Observatory, which is responsible for its operation under the auspices of the Chinese Academy of Sciences. The TMRT saw its first light observing the massive star forming region W3(OH) in 2012. The TMRT features a modified Cassegrain antenna with primary focal ratio of 0.32, and a 6.5-m sub-reflector. The primary reflector consists of 14 rings for a total of 1008 high-precision solid panels, while the sub-reflector consists of 3 rings with 25 double-layered, aluminum honeycomb reflecting panels. A total of 1104 actuators were installed at the junctions of the primary reflector panels with the antenna backbone structure so as to compensate for gravitational deformations in the primary reflecting surface during tracking.

The TMRT is characterized by high operating frequencies and wide receiving bandwidths. The telescope currently has 7 cryogenic receivers covering eight frequency bands in the range 1-50 GHz, i.e., L-band (1.25-1.75 GHz), S/X-dual band (2.2-2.4/8.2-9.0 GHz), C-band (4-8 GHz), Ku-band (12-18 GHz), X/Ka-dual band (8-9/30-34 GHz), K-band (18.0-26.5 GHz), and Q-band (35-50 GHz). Observations with the L-band receiver are done by tilting the sub-reflector towards the phase center of the feed. All the other receivers are mounted on a mechanically rotating turret able to place the selected feed horn at the secondary focus. The receiver change process can be accomplished within one minute.

The main TMRT back-end is a digital system, DIBAS, for single-dish observations. It is an updated version of the NRAO-VEGAS, customized with the addition of NRAO-GUPPI (Green Bank Ultimate Pulsar Processing Instrument) modes. It can support both spectral line and pulsar observing modes. For molecular line observations, DIBAS supports 29 different observing modes, including 19 single sub-band modes and 10 eight sub-band modes. The bandwidth varies from 1250 to 11.7 MHz for the single sub-band modes 1-19. The spectrometer supports eight fully tunable sub-bands within a 1300 MHz bandwidth. The bandwidth of each sub-band is 23.4 MHz for modes 20−24 and 15.625 MHz for modes 25−29. For pulsar observations, DIBAS supports both pulsar searching and online folding mode. Both coherent and incoherent de-dispersion observation modes are supported. Additional back-ends for VLBI and for other types of observations are available.

TMRT control system supports standard position switching, gridding, on-the-fly mapping (($\alpha$,$\delta$), (Az,El) and (l,b) coordinate systems) and various VLBI observing modes.

### 2.4. Parkes Radio Telescope (Australia)

The CSIRO Parkes Radio Telescope (Robertson 1992) is a 64-m parabolic antenna operating as part of the Australia Telescope National Facility (ATNF). It has a focus cabin that can accommodate two translatable receiver packages.

The current receiver fleet covers the range 700 MHz to ~25 GHz, with 9 cryogenic receiver packages: the '10/50' (700-764 MHz/2.6-3.6 GHz); the 13-beam receiver (1.2-1.5 GHz); H-OH (1.2-1.8 GHz); GALILEO (2.1-2.5 GHz); METHANOL6 (5.9-6.8 GHz); MARS (8.1-8.5 GHz); S/X,C (2.2-2.5/8.1-8.7, 4.5-5.1 GHz); Ku (12-15 GHz); and the K/13mm (16-26 GHz). These receivers are currently being consolidated into one forthcoming single installation of multiple UWB receivers to cover the same frequency range. The first, Ultra-Wideband Low (UWL), will operate from 700 MHz to 4 GHz with a system temperature of approximately 22 K and was commissioned in 2018. The complementary wideband receiver(s) to cover the band from 4 to 25 GHz are in the design phase and would share the same cryogenic package. In addition, a cryogenically cooled PAF is also under design, with an uncooled prototype tested on telescope in 2016. Plans are that this PAF will share the focus cabin with the wideband receivers to provide full frequency and survey capabilities with no receiver installation changes.

In terms of back-ends available at PARKES there is the FPGA based Digital Filterbank (DFB), which provides time and spectral domain capabilities, the HIPSR and BPSR Graphical Processor Unit (GPU) based systems (serving multi-beam spectral line and time domain astronomy), and the standard suite of VLBI back-ends including DAS. A new GPU-based back-end, PKBE, is under development to serve the UWL and all future back-end needs. Currently, there is also a back-end system installed by the University of California, Berkeley, to serve the Breakthrough Listen project.

Observing modes include the full range of those expected by the astronomical community including position switching, frequency switching, on-the-fly mapping, pointed, and gridding. Observations may be performed from any location served by the internet, with the telescope itself being unattended for part or all of the time (Edwards et al. 2014).

### 2.5. Sardinia Radio Telescope (Italy)

The Sardinia Radio Telescope is a 64-m diameter Gregorian reflector inaugurated in 2013 (Bolli et al. 2015). Besides the classical primary and secondary focus, four foci are also available in beam-waveguide mode. The antenna can operate in the frequency range 0.3 to 116 GHz thanks to the active surface, which allows the reconfiguration of the ideal shape of the primary mirror at any elevation.

The SRT is currently equipped with dual-polarization cryogenic receivers with bandwidths 305-410 MHz, 1.3-1.8 GHz, 5.7-7.7 GHz, and 18-26 GHz, plus two room-temperature receivers at 8.2-8.6 GHz and 31.8-32.3 GHz. All the receivers are single-feed, except for the 18-26 GHz seven-feed receiver and the 8.2-8.6/31.8-32.3 GHz coaxial-feed, dual-frequency one. Currently, three new front-ends are under development to cover the frequency ranges 3.0-4.5 GHz (multi-feed), 4.2-5.6 GHz (single-feed), and 33-50 GHz (multi-feed). Frequency-converted signals are routed to the back-ends located in the shielded antenna control room via fiber optics.

Up to 2 GHz bandwidth per polarization can be processed by either a fourteen-input analog total power back-end or one of the two spectro-polarimeters available (either a fourteen-input with 125 MHz bandwidth and 2048 channels or a two-input with 1.8 GHz bandwidth and 16384 channels). A DFB correlator is mainly used for pulsar observations with 1 GHz bandwidth, two inputs, and up to 8192 channels. A DBBC is also available, working with MK5 or Flexbuff recording systems for VLBI observations. The use of the MK5 implies the shipping of disk packs to the correlator while the new Flexbuff system sends the data to the correlator in real time via fiber optics.

The configurable control system supports standard observing strategies, such as position switching, pointed observations, raster scans, and on-the-fly mapping. The telescope can be operated both remotely and locally.

### 2.6. Nobeyama Radio Telescope (Japan)

The Nobeyama 45-m Telescope (Ukita & Tsuboi 1994) is a 45-m diameter Cassegrain-modified-Coude reflector radio telescope that has been operated by the Nobeyama Radio Observatory (NRO), a branch of the National Astronomical Observatory of Japan (NAOJ), National Institutes of Natural Sciences (NINS), since 1982. The telescope is operated in the frequency range of ~20 - 116 GHz (wavelengths ~15 mm - 2.6 mm) at a highland of 1350 m altitude in Nagano, Japan. The Nobeyama 45-m Radio Telescope was certified as an "IEEE Milestone" in 2017. The main reflector of the telescope consists of ~600 panels supported by a steel truss structure, which is also covered by backside insulator panels. This backup structure and the appropriate positioning and tilt of the sub-reflector realize the homologous deformation. The telescope adopts an Alt-Az mount system and a master-collimator to achieve precise pointing accuracy. The radio signal collected with the main- and sub- reflectors is guided to the receiver cabin with the common beam-guide optics, which consists of a plane mirror, a pair of ellipsoidal mirrors, and a rotatable plane mirror. The successive beam-guide optics in the receiver cabin guides the radio signal to a selected receiver among the 11 receiver ports, and also supports chopper-wheel calibrators.

In the receiver cabin, five heterodyne receivers are now accommodated, that is, H22 (20-25 GHz, single-beam, dual-circular polarization), H40 (42.5-44.5 GHz, single-beam, left circular polarization), Z45 (42-46 GHz, single-beam, dual-linear polarization; Nakamura et al. 2015), T70 (71.5-92.0 GHz, single-beam, dual-linear polarization), and FOREST (80-116 GHz, four-beam, dual-linear polarization; Minamidani et al. 2016). T70 and FOREST are sideband-separating SIS receivers.

The back-end system for spectral line observations is the combination of a 3-bit and 4 Gsps digitizer, OCTAD-A (Oyama et al. 2012), and a FX-type correlator, SAM45 (Kuno et al. 2011), which is the same as a subset of the ALMA ACA Correlator (Kamazaki et al. 2012). This back-end can process sixteen 2-4 GHz signals simultaneously. Back-end systems for continuum observations, and for VLBI observations are also provided.

The telescope control system, COSMOS-3 (Morita et al. 2003), controls the antenna, optics, receivers, and back-ends, in accordance with observing scripts generated by dedicated GUI software. This system supports observation modes such as the single point, on-the-fly (Sawada et al. 2008), cross point, multi points, On-On (Nakajima et al. 2013) observations etc. Remote observations are possible via VNC connections.

### 2.7. Yebes Radio Telescope (Spain)

The Yebes 40-m radio telescope (Bachiller et al. 2007), operated by the Instituto Geográfico Nacional of Spain, performed its first-light observation in 2007. The telescope is a Nasmyth-Cassegrain turning-head antenna and observes in the range 2-116 GHz.

It is equipped with several cryogenic receivers at 2.20-2.37 GHz, 3.22-3.39 GHz, 4.56-5.06 GHz, 5.9-6.9 GHz, 8.18-8.98 GHz, 20-26 GHz, 41-50 GHz, and 83-116 GHz, plus one room-temperature Ku-band receiver mounted at the prime focus position for holography measurements. Currently, simultaneous observations in K/Q band are available, each with 2.5 GHz instantaneous bandwidth.  Simultaneous K/Q/W band observations will be implemented by 2018 in the framework of the NanoCosmos ERC Synergy project, when the Q-band and W-band receiver bandwidths will be enlarged to 31.5-50.0 GHz and 72-90.5 GHz, respectively. Also the instantaneous bandwidth will be upgraded to 18.5 GHz for both front-ends so that the full Q and W bands will be measured in a single scan.

Several spectral back-ends are connected to the Intermediate Frequency (IF): a wide-band Fast Fourier Spectrometer (16 modules, 2.5 GHz bandwidth each) and a six-channel continuum detector (2 GHz bandwidth). Currently, the total instantaneous bandwidth is 9 GHz at 45 GHz. The telescope is also equipped with VLBI back-ends: three DBBC and two Flexbuff systems, which act as recorders and are connected at 10 Gbps speed to the Internet via RedIris, allowing real time 4 Gbps transfers to any correlator in the world.

As a result, the telescope can be used in single-dish and astronomical (EVN, GMVA, Radioastron) and geodetic VLBI. Its available observing modes are position switching and frequency switching. On-the-fly and grid maps are available too. Continuum observations are available for pointing a focus calibration only. In addition, pointing and focus calibration may optionally be performed as pseudo-continuum observations. Observations are performed by antenna operators using macros and procedures defined and prepared by the observer with the help of YEBES support astronomers.

### 2.8.  Medicina and Noto Radio Telescopes (Italy)

The Medicina and the Noto radio telescopes are 32-m diameter Cassegrain reflectors currently operating in the frequency range 1.4 to 26 GHz and 1.4 to 43 GHz, respectively. The two radio telescopes are identical in their structure, however NOTO is equipped with an active surface composed of 240 actuators enabling high-frequency capabilities up to 86 GHz.

MED is equipped with dual-polarization cryogenic receivers at 2.20-2.36/8.18-8.98 GHz (coaxial-feed dual-frequency), 4.3-5.8 GHz and 18-26 GHz (dual-feed), plus three room-temperature front-ends at 1.35-1.45 GHz, 1.59-1.72 GHz, and 5.9-7.1 GHz. NOTO is equipped with dual-polarization cryogenic receivers at 4.6-5.1 GHz, 21.5-23.0 GHz, and 39.0-43.5 GHz, plus three room temperature receiving bands at 2.20-2.36/8.18-8.58 GHz (coaxial dual frequency) and 5.1-7.2 GHz. For MED and NOTO, receivers may be placed at both the primary and secondary focus positions.

Frequency-converted signals are routed via fiber optics (MED) or coaxial cables (NOTO) to the antenna control room, where a distributor broadcasts the signals to the back-ends. Up to 2 GHz bandwidth per polarization can be processed at both stations. A four input analog total power back-end plus two spectro-polarimeters identical to those available at the SRT are installed at MED, while a two input analog total power back-end is available at NOTO. Both telescopes are equipped with a DBBC plus a MK5 or Flexbuff recording system for VLBI observations and are connected at 10 Gbps to the Internet.

MED and NOTO share the same configurable observation control system (common to the SRT as well) supporting standard observing strategies such as position switching, pointed observations, raster scans, and on-the-fly mapping. The telescopes can be operated both locally and remotely via VNC connection.

### 2.9.  Pico Veleta Radio Telescope (Spain)

The 30-m Pico Veleta telescope (Baars et.al. 1987, 1994) is one of the two observatories operated by IRAM (Institut de Radioastronomie Millimétrique). It is located in the south of Spain, in Sierra Nevada, at an altitude of 2850 meters and had its first light in May 1984. The 30-m telescope is a Cassegrain type antenna with a semi-homologous mechanical design. The structure of the main parabolic dish is completely enclosed to shield it from weather influences, and its temperature is carefully controlled to keep thermal deformations to a minimum. The low RMS surface error of the primary (~60 μm) allows observations in the entire millimeter band.

The telescope is currently equipped with three different front-ends that cover the frequency range from 70 to 370 GHz with very small gaps, closely matching the main atmospheric transmission windows: EMIR is a four-band (3 mm, 2 mm, 1.3 mm and 0.8 mm), single-pixel, dual-sideband, 8-GHz-wide IF, dual-polarization heterodyne receiver; HERA is a 3 x 3 multi-beam with dual-polarization, single-sideband receiver operating in the 1.3 mm band; and the recently installed NIKA2 is a continuum camera based on LEKID detectors with a 616 pixel array in the 2 mm band plus another 2 x 1140 pixel array in the 1.3 mm band. With the help of a warm rotating half-wave plate located on the receiver input, the two 1.3 mm band orthogonal arrays can be combined to make a very sensitive polarimeter for continuum observations. Two multi-beam receivers are under development: a 5 x 5 pixel array in the 3 mm band, with dual-polarization mixers and 8 GHz IF bandwidth; and a 7 x 7 pixel array in the 1.3 mm band with identical mixer specifications.

With the existing FTS spectrometers, a maximum of 32 GHz (4 complete IF bands, 8 GHz wide each) of instantaneous bandwidth can be analyzed with a channel width close to 200 kHz. Higher resolution on a narrower frequency band is obtainable by changing the FTS resolution to 50 kHz. Alternatively, the VESPA spectrometer can be used. This auto-/cross-correlation system offers variable channel widths in the 3.3 kHz to 2.5 MHz range. A third spectrometer, WILMA, also based on an auto-correlation scheme, offers up to 18 independent blocks of 1-GHz-wide band and 2 MHz resolution. VESPA, when coupled to the EMIR receiver, is also the base of a continuum and spectral polarimeter where all four Stokes parameters can be measured on any of the four EMIR bands.

Apart from the standard preparation and calibration modes, like pointing, focusing, calibration of receivers and atmospheric parameters, the control program (NCS) of the 30-m telescope offers the following observing modes: total power, beam switching, position switching, Wobbler switching, frequency switching, on-the-fly mapping and polarimetry modes. Remote observations are routinely possible, although they are only recommended for short observations and restricted to experienced 30-m observers.

### 2.10.    20m and 25m Onsala Radio Telescopes (Sweden)

The Onsala Space Observatory (OSO) is the Swedish National Facility for Radio Astronomy, hosted by the Department of Space, Earth and Environment at Chalmers University of Technology in Göteborg and operated on behalf of the Swedish Research Council. OSO operates five telescopes at Onsala, a 25-m diameter cm-wave telescope, a 20-m diameter mm-wave telescope, a LOFAR station, and two 13-m antennas for geodetic VLBI. OSO is involved in both cm-wavelength VLBI, through their participation in the EVN and in mm-wavelength VLBI via the Global Millimetre VLBI Array (GMVA).

The 25-m diameter polar-mounted decimetre-wave telescope was built in 1963. ONS25 is equipped with receivers for 1.2-1.8 GHz, 4.5-5.3 GHz and 6.0-6.7 GHz. VLBI observations make use of a DBBC back-end and data are recorded on a Flexbuff system. ONS25 is currently only used for VLBI science observations. Pointing and calibration (both continuum and spectral line) observations are performed using a GNU Radio-based FFTS spectrometer.

The 20-m diameter elevation-azimuth-mounted millimeter-wave telescope is of Schmidt-Cassegrain type. It is enclosed by a radome, which was fully upgraded in the summer of 2014. The telescope itself was built in 1975-76 and upgraded in 1992. Front-ends and back-ends are updated regularly. Presently, ONS20 is equipped with receivers at 2.2-2.4 GHz, 8.2-8.4 GHz, 18-50 GHz, 67-87 GHz (4 mm), and 85-116 GHz (3 mm). The 3 mm (Belitsky et al. 2015) and 4 mm receivers were installed in March 2014 and October 2015, respectively. Spectral line observations can be accomplished with FFTS back-end spectrometers. The OSO20 telescope provides three observing modes: beam switching, position switching, and frequency switching. The ONS20 telescope is equipped with a VLBI system identical to the ONS25 system, allowing OSO to routinely perform VLBI observations with both telescopes simultaneously.

Both ONS20 and ONS25 can be operated remotely; there is a plan to introduce dynamic scheduling on ONS20.

### 2.11.    Mopra Radio Telescope (Australia)

The Mopra 22-m antenna is part of the Australia Telescope National Facility operated by the CSIRO. A detailed characterization of the antenna in the mm-bands has been essential for the observing programs and is described in (Ladd et al. 2005; Urquhart et al. 2010; Foster et al. 2013). From October 2012, the Mopra antenna is no longer fully supported for National Facility observations except for VLBI. The telescope is available to groups outside CSIRO that can provide funding for telescope maintenance and operations, with these groups being provided with guaranteed observing time in return. MOPRA has operated successfully in this mode and is also potentially available for any other uses. A severe bushfire in 2013 resulted in significant damage to the on-site building but the telescope itself and the key data-taking equipment were spared by the fire. The telescope was returned to operations in May 2013.

The antenna can operate in the range 1.2-117 GHz and is currently equipped with systems at: 1.2-3.0 GHz; 4.4-6.7 GHz and 8.0-9.2 GHz; 16-27 GHz; 30-50 GHz; and 76-117 GHz. MOPRA has been a key element of the Australian Long Baseline Array (LBA) especially at cm-wavelengths and played a critical role for the VSOP satellite VLBI observing program. It is also used for single-dish observations, particularly in the millimetre bands.

The Mopra Spectrometer (MOPS) back-end system processes two 8-GHz-bandwidth IFs for each polarization and is capable of observing several spectral lines simultaneously.

MOPRA observers must carry out their own observations and full remote observing from anywhere in the world is available for suitably qualified observers.

### 2.12.   Korean VLBI Network (Korea)

The Korean VLBI Network (Kim et al. 2004) consists of three 21-m shaped Cassegrain reflectors located in Seoul (Yonsei), Ulsan (Ulsan), and Jeju Island (Tamna) and is operated by the Korea Astronomy and Space Science Institute (KASI). The baseline lengths range from 305 to 476 km. It is a facility dedicated to VLBI and focused on mm-wavelength observations. Dish surface adjustment, which will enable participation in international campaigns such as the Event Horizon Telescope (EHT) at 230 GHz, is under discussion. The construction of three new telescopes, that will in particular increase the number of short baselines, is in the planning phase and the completion is foreseen in 2023. The KASI hosts the Array Operation Center in the Daejoen headquarters. All the three observatories and the KASI headquarters are inter-linked with a wide-band research network called Kreonet supporting observation data transfer and remote operation from any of the four locations.

The VLBI mode is operational in four frequency bands: 21-23 GHz; 42-44 GHz; 85-95 GHz; and 125-142 GHz, all in dual polarization. The KVN Yonsei antenna has a wider coverage in the K band, namely 18-24 GHz. The KVN Ulsan telescope has an additional wide-band receiver covering 6.4 to 9.0 GHz. All the front-ends are cryogenic single-feed systems. A key feature of the KVN is simultaneous multi-frequency observations using adaptive optics (Han et al. 2008). Any combination of the available frequency bands is possible (see scientific cases for some examples). Data are sampled in the antenna receiver cabins and transmitted to the observatories through optical fiber.

Currently up to 8 Gbps data recording, 2-GHz bandwidth with 2-bit sampling is operational. Mark 6 recorders are in use. The KASI has two correlators, a DiFX correlator and a hardware correlator in the Korea-Japan Joint Correlator Center (KJCC) at Daejeon. KJCC provides flexible data format conversion to support various types of international cooperation. The introduction of a wider-bandwidth back-end and a recording system up to 64 Gbps is underway.

Besides VLBI mode, complementary single-dish observing modes, e.g. position switching, five-points and cross scan, are also available in the same bands.

### 2.13.    VLBI Exploration of Radio Astrometry (Japan)

VLBI Exploration of Radio Astrometry is a VLBI array dedicated to radio astrometry, being operated by the National Astronomical Observatory of Japan (NAOJ) in collaboration with universities in Japan (e.g., Kobayashi et al. 2003). The VERA array consists of four identical 20-m radio telescopes located at Mizusawa, Iriki, Ogasawara, and Ishigaki-jima, with baseline lengths ranging from 1000 to 2300 km.   Array operation is carried out remotely from the Array Operation Center located at Mizusawa. VERA antennas have a standard Cassegrain-type dish on an Alt-Az mount, with minimum and maximum elevation limits of 5 and 85 degrees. The most unique feature of the telescopes is the "dual-beam" system (Kawaguchi et al. 2000), which allows the simultaneous observation of two adjacent sources on sky. This allows accurate measurement of the relative position of the two sources by cancelling out tropospheric phase fluctuations. The dual-beam system consists of two Stewart-mount platforms installed in the Cassegrain-focus receiver cabin. The two mounts are equipped with receivers, which are steerable along the focal plane, so that the separation angle between the two beams can be changed between 0.5 and 2.2 deg.

The main receiving bands for VERA are 22 and 43 GHz, for which cooled single-pixel receivers are installed on the two platforms for dual-beam observations of $H_2O$ and SiO masers as well as continuum sources. In addition, a room-temperature, single-beam, single-pixel C-band receiver is installed for observations of methanol ($CH_3OH$) masers. Single-pixel, single-beam S and X receivers working at room temperature are available for geodetic observations.

As for back-ends, each VERA station has a digital data acquisition system to record the time series of the received voltage. The current sampling rate is 2 Gbps per beam, which provides a maximum recording bandwidth of 512 MHz with a 2-bit quantization. Sampled data are recorded onto OCTADISK hard-disk-drive recorders, and the disk packs are shipped to the correlation center located at the Mizusawa campus of NAOJ for further processing.

## 3.    Science highlights

In this section we aim at highlighting the large variety of astrophysical studies that are conducted with the radio astronomical instrumentation at the various facilities considered in this survey. As a general consideration, the scientific usage of a receiver depends not only on its characteristics, like frequency and number of feeds, but also on the telescope on which it is mounted, the available back-ends and the type of observation to be conducted (single-dish or interferometry). The analysis of the scientific usage of these international facilities shows that single-dish observing programs typically use receivers at the higher frequencies, like the C, X-, K, Q and W bands. Lower frequency front-ends are characterized by very large beam sizes and are generally used for VLBI or geodetic observation, with single-dish usage mostly limited to the larger antennas. The versatility of new digital back-ends nowadays available at the majority of the radio Observatories allows for instance the implementation of RFI mitigation strategies, nowadays  becoming a requirement to conduct successful single-dish science at many sites. The advent of new, dense multi-feed front-ends has dramatically increased the efficiency in mapping large areas of the sky, boosting the survey capabilities of single antennas. The same receiver design, if adopted at high-frequency, also increases the scientific performance for both continuum and spectral line studies of astrophysical sources by permitting the correction of atmospheric effects within the same observation. Almost all the considered radio telescopes regularly participate in international networks like the VLBI, the international VLBI service for Astrometry and Geodesy (IVS), and GMVA to study science cases ranging from geodesy and Earth parameters to the physics of the most distant objects in the Universe and cosmological probes. This translates in the need for a common instrumental equipment among the participating stations as well as a strong level of coordination and exchange of experience and information.

A further significant improvement is expected in the near future thanks to new technological developments in the field of PAF, UWB and simultaneous frequency front-ends. Despite their current development only at moderate frequencies, PAF technology offers improved sensitivity and fast mapping capabilities for both single-dish telescopes and interferometers, with astrophysical applications including survey studies of galaxies up to high redshift and the search of transient sources. Thanks to their continuous frequency coverage, UWB receivers like the BRAND one under design for EVN and IVS applications, will be particularly efficient in multi-wavelength mapping and will allow simultaneous spectroscopic and polarimetric studies of the physical conditions of various astronomical objects. Despite being mostly designed for VLBI and geodetic applications, thanks to their large bandwidths these front-ends could have useful application also in classical single-dish programs like long term flux monitoring of sources. Wide-band, multi-wavelengths receivers are recognized as essential elements for a leap forward in the 20-100 GHz domain. In recent years, technical progress has also opened up the possibility of high-frequency simultaneous observations (e.g Rioja et al 2015) in particular for mm-VLBI studies of compact radio emitting regions inaccessible at lower frequencies and complex molecular species. Such developments also increase the synergy among telescope networks like for instance the GMVA and ALMA.

The science topics that will be illustrated in the following sub-sections are not intended to be an exhaustive list of all the possible scientific applications, the aim being to give an overview of the main astronomical topics that are currently driving single-dish and VLBI observations, both for continuum and spectroscopic techniques, at the various radio telescopes.

### 3.1. Green Bank Telescope

The GBT has a broad science program that crosses many scientific disciplines. It is used as the receiving element for bi-static radar studies of the Moon, Planets, and near earth asteroids (Campbell 2016; Margot et al. 2007; Asteroid 1999 JD6, http://www.jpl.nasa.gov/spaceimages/details.php?id=PIA19647).

Chemical studies of interstellar clouds led to the discovery of numerous pre-biotic molecules including the first known interstellar anion, and the first interstellar chiral molecule (McCarthy et al. 2006; McGuire et al. 2016). The GBT detected more than 1000 new Galactic HII regions (Anderson et al. 2015) and revealed mm-sized interstellar dust grains in Orion (Schnee et al. 2014). Star-forming cores and the evolution of turbulence during cloud collapse have been investigated by mapping molecular clouds in Ammonia and other species (Friesen et al. 2017; Seo et al. 2015).

Studies of the 21cm line are done in the Milky Way and other galaxies. HI cloud systems were discovered around nearby galaxies (Thilker et al. 2004; Wolfe et al. 2013) and the intensity mapping technique developed at GBT allowed the determination of the HI content of intermediate redshift galaxies (Chang et al. 2010). $H_2O$ maser studies in accretion disks around nuclear black holes in galaxies yielded independent and precise measurements of the black hole mass and the distance of the galaxies (Braatz et al. 2013) while highly redshifted molecular lines have been used to constrain models of the formation and evolution of massive galaxies at high redshift (Negrello et al. 2010).

The GBT has discovered the lowest-frequency Fast Radio Burst (FRB) to date and measured its polarization (Masui et al. 2015). It has also provided very high angular resolution measurements of the Sunyaev-Zel'dovich effect in galaxy clusters (Korngut et al. 2011) and data from the GBT contributed to studies of the structure of the local Universe (Tully et al. 2014) and the baryon content of low mass dwarf galaxies (Geha et al. 2006).

Pulsars and other compact objects are studied with the GBT, which is a key instrument in the North American Nanohertz Observatory for Gravitational Waves (Demorest et al. 2013, Arzoumanian et al. 2016). The GBT has been used to discover the fastest pulsar and the most massive pulsar, as well as more than two dozen pulsars in the globular cluster Terzan 5, and to provide stringent tests of the theory of general relativity in the strong field case (Demorest et al. 2010; Prager et al. 2017; Breton et al. 2008).

When used as an element of a Very Long Baseline Interferometer, the GBT's added sensitivity has allowed sub-milliarcsecond resolution measurement of the radio jet in the galaxy M87 (Hada et al. 2017). GBT data was also used to place limits on the variations in the gravitational constant, G, several kpc from the Solar System, and on violations of the equivalence principle (Zhu et al. 2015; Ransom et al. 2014).

### 3.2. Effelsberg Radio Telescope

The large number of receivers together with several specialized back-ends dedicated to different observing modes make EFF a telescope well suited for a variety of single-dish scientific applications as well as the participation in a number of interferometric networks.

Among the recent scientific results are survey observations like the Effelsberg-Bonn HI Survey (EBHIS) whose first data release was in 2016 (Winkel et al. 2016). EBHIS was the first large-scale survey of neutral hydrogen conducted with a 100-m-class telescope and delivers unprecedented data in terms of sensitivity and resolution. The telescope is used for molecular line surveys, e.g. towards Orion-KL at 1.3cm (Gong et al. 2015). Observations of methanol absorption lines towards the lensing galaxy PKS1830-211 allowed a limit to be placed on the possible cosmological variations of the electron-to-proton mass in the early Universe (Bagdonaite et al. 2013).

The Effelsberg telescope is used in a number of pulsar observations, (see e.g. Desvignes et al. 2016). Sensitive broad-band measurements of a magnetar in the vicinity of the black hole in the center of our Milky Way allowed the detection of extremely high RM values (Eatough et al. 2013). For more than 20 years the Effelsberg telescope has been performing monthly timing multi-frequency observations within the European Pulsar Timing Array (EPTA). Furthermore, together with the Sardinia Radio Telescope, the Jodrell Bank telescope, the Nancy Radio Telescope, and - initially - the Westerbork Synthesis Radio Telescope, it is part of the Large European Array of Pulsars (LEAP). LEAP realizes a virtual telescope with an effective aperture equivalent to a 195-m diameter circular dish (Bassa et al. 2016).

There is a long history of on-the-fly continuum observations at EFF, with the first studies of M31 started in early days of the telescope. Several deep observations of the extended emission of the Andromeda Galaxy have been performed at various wavelengths (see e.g. Giessübel & Beck 2014). Due to the high frequency agility and very good polarization properties of the receivers in the secondary focus, the telescope is well suited for broad-band Active Galactic Nuclei (AGN) flux density and polarization monitoring, as demonstrated e.g. by Fuhrmann et al. (2016).

The 100-m telescope participates regularly in VLBI observations within the EVN, also in real-time mode (eVLBI) making use of data transfer for immediate correlation, and observes frequently as part of the High Sensitivity Array (with the GBT, VLBA, VLA, and Arecibo) and with the RadioAstron satellite telescope (see for instance Gomez et al. 2016).

Twice a year the 100-m participates in observations of the GMVA at 3nmm wavelength, allowing micro-arcsecond resolution studies of the structure of radio galaxies down to few hundreds of Schwarzschild radii from the central black hole (see e.g Boccardi et al. 2016). Finally, EFF is member of IVS and performs several geodetic VLBI observing runs per year.

### 3.3. TianMa Radio Telescope

With its good sensitivity and wide frequency coverage, the TMRT as a single-dish plays an important role in the astronomical observation and research of molecular spectral lines as well as the study of pulsars. The TMRT is also used for observations of specific targets including radio blazars, microquasars, and X-ray binaries, as well as for continuum monitoring observations. Research activities on the fast time variation of AGN ()and the transient phenomenon of X-ray binaries are key topics for high-sensitivity single-dish observations, when combined with other means of observation.

The TMRT is a key element of the VLBI network in China and in the world, and significantly increases its resolution and capability for VLBI observations. It is China's first radio telescope capable of observing at the 7-mm wavelength, opening up new areas of millimeter VLBI observation. As a VLBI station, it has joined observations with most of the world-wide VLBI networks, including CVN (Chinese VLBI Network), EVN, VLBA, LBA, IVS, East Asian VLBI Network, and space VLBI observations with RadioAstron. In particular, a test observation of the TMRT with the dedicated high-frequency VLBI network KaVA (KVN and VERA Array) composed of 3-station KVN and 4-station VERA was successfully performed, demonstrating the promise of 43 GHz VLBI observations with the TMRT in the future.

### 3.4. Parkes Radio Telescope

The Parkes telescope has a varied science output, which has been driven by continual evolution of the facility and its front- and back-end technologies. Parkes was the first telescope to detect a FRB (Lorimer et al. 2007), and is currently the world leader with the detection of 20 out of 25 the published FRBs (Thornton et al. 2013, Keane et al. 2016, Keane et al. 2018). Parkes has been a pulsar survey machine with more than one half of the currently known pulsars detected with the telescope, and is the key instrument of the Parkes Pulsar Timing Array (Manchester et al. 2013) with 10 years of high precision timing data providing stringent limits on long wavelength gravitational waves (Shannon et al. 2015).

In addition to time domain science, PARKES has been very active in various atomic and molecular line surveys, establishing significant data sets with legacy value including the Galactic All Sky Survey (GASS, McClure-Griffiths et al. 2009), the Methanol Multibeam Survey (MMB, Green et al. 2009), HI Parkes All-Sky Survey (HIPASS, Staveley-Smith et al. 1996) and several others. The significant work done with atomic hydrogen, and the pulsar results, were both facilitated by the prolific multi-beam receiver. Recently it provided confirmation of the first detection of a Chiral molecule outside our solar system (McGuire et al. 2016) and has had an active role in SETI, including Project Phoenix (Tarter et al. 1997) and more recently with the Breakthrough Listen initiative (Isaacson et al. 2017). Parkes is also a crucial component of the southern hemisphere VLBI network, the Long Baseline Array, through which it is contributing to the southern hemisphere equivalent of Bar and Spiral Structure Legacy (BeSSeL, Brunthaler et al. 2011) Survey.

### 3.5. Sardinia Radio Telescope

The key science that can be addressed with the SRT, also through VLBI observations with the EVN and Radioastron, include: gravitational wave detection experiments and pulsar studies; transient sources studies; galactic and extragalactic surveys at high frequency; high resolution spectroscopy; and space applications. A more detailed illustration of the SRT science highlights can be found in Prandoni et al. (2017).

Early Science operations started in 2016 demonstrate the telescope capabilities for both continuum and spectro-polarimetric studies. Together with interferometric data, SRT observations allowed the detection of diffuse radio emission associated with a large-scale filament of the cosmic web, indicative of the emergence of a very faint and new population of radio sources (Vacca et al. 2018). Multi-wavelength studies of radio galaxies and relics in clusters are used as a probe of the intra-cluster magnetic field power spectrum and the cluster merging history (Govoni et al. 2017, Loi et al. 2017). The physics of supernova remnants and their interaction with the interstellar medium have been studied through sensitive, high-resolution observations to build the integrated and spatially resolved spectra at various frequencies (Loru et al. 2019, Egron et al. 2017).

The SRT participates in observations of the EVN VLBI network and participates in RadioAstron observations. Furthermore, together with other European single-dish facilities, the SRT is part of the LEAP project dedicated to monthly monitoring of millisecond pulsars.

### 3.6. Nobeyama Radio Telescope

The Nobeyama 45-m Telescope provides various capabilities, such as large-area mapping, simultaneous multi-line observations and deep integration observations toward point-like sources, and performs many observations with various molecular lines toward various targets mainly based on the "Open Use" framework.

In addition, Nobeyama Radio Observatory (NRO) Legacy Projects are launched occasionally and spearhead new scientific observations with new instruments and capabilities. Recent outcomes from the NRO Legacy Projects are a CO(J=1-0) multi-line survey toward the Galactic plane (Umemoto et al. 2017); CO(J=1-0) mapping toward nearby galaxies (e.g., Muraoka et al. 2016); and studies toward some star-forming regions in the Milky Way (e.g., Kong et al. 2018). Data for these projects are open to the community through the Japanese Virtual Observatory (JVO; http://jvo.nao.ac.jp/portal/v2/).

Finally, NOB is also used as an element of various VLBI networks.

### 3.7. Yebes Radio Telescope

The Yebes 40-m radio telescope is part of the following international networks: EVN; GMVA; IVS; and offers time to RadioAstron, the VLBI satellite orbiting around the Earth. It is also open to national and foreign groups who apply directly to the Yebes Observatory for observing time. A large fraction of single-dish observations requested by external investigators are dedicated to the Nanocosmos project. YEBES is one of the most relevant telescopes in the EVN because of the size of its main dish, the frequency coverage, and the low system temperature of its receivers. YEBES becomes the reference antenna of the EVN, particularly when EFF is not available. The 40-m is also a key element in the GMVA, formed by five European telescopes and some VLBA telescopes from the USA, and provides information of maser sources intensities prior to each observing session. Similarly to other radio telescopes, YEBES is also an important element of the IVS contributing for instance to the determination of the Earth Orientation Parameters solution from the different analysis centers across the world.

### 3.8.  Medicina and Noto telescopes

The Medicina and Noto 32-m telescopes regularly take part in EVN experiments as well as in RadioAstron space VLBI, and collaborations exist with international arrays like KVN and VERA. NOTO also participated in GMVA experiments at 43 GHz. Interferometric science topics mainly focus on: high resolution studies of radio sources and jets; the mechanisms driving the formation and evolution of AGN; and spectroscopic and polarimetric studies of pulsars and of maser sources in star forming regions (e.g. Giovannini et al. 2018; Schulz et al. 2018; Deane et al. 2014; Du et al. 2014; Moscadelli et al. 2018). Both MED and NOTO participate on a monthly basis in IVS observations, contributing to the definition of the Celestial and Terrestrial Reference Frames and to the determination of the Earth parameters (Altamimi et al. 2016, Madzak et al. 2016).

In the last ten years MED antenna (and more recently NOTO) experienced a significant increase of single-dish activity thanks to the advent of new generation instrumentation as well as a state-of-the-art observing software allowing fast on-the-fly observations. Radio continuum monitoring of sources complementary to international projects at other wavelengths like Planck, FERMI and AGILE-GLAST (e.g. Procopio et al. 2011; Orienti et al. 2016; Carnerero et al. 2015) is regularly conducted, with  the formation and evolution of galaxies and AGN, and the physics of radio sources as the main science topics (Vercellone et al. 2019; Ackermann et al. 2014; Ceglowski et al. 2015). Large-area extragalactic surveys are also a science case well suited for these medium-size antennas (e.g. Righini et al. 2012). Long-term spectroscopic observations have been regularly conducted at MED for many decades (Felli et al. 2007). The recent availability of modern digital spectro-polarimeters is reactivating the interest for molecular studies of star forming regions, stellar evolution and interstellar medium properties (e.g. Lekht et al. 2018).

### 3.9.  Pico Veleta Radio Telescope

The 30-m telescope has been used to investigate a wide range of topics (from planetary atmospheres and comets to stellar atmospheres, galactic and extragalactic star forming regions to molecular and thermal emission of high redshift galaxies) and physical and astronomical processes. About half of the observations granted at the observatory target Galactic topics, the other half nearby galaxies, galaxies at high-redshift, and cosmology research.

Molecular spectroscopy is a preferred research field of IRAM30. The broadband frequency coverage in the millimeter band helps performing surveys in almost the entire millimeter band, where a great number of molecular species have been identified for the first time.

Thanks to the dedicated VLBI instrumentation, the observatory regularly participates twice per year in GMVA observations at 3 mm wavelength. The 30 m telescope also participates in 1 mm observations within the framework of the EHT. One of its key goals is to image the immediate surroundings of the supermassive black hole in the center of the Milky Way**.**

### 3.10.    Onsala 20 m and 25 m Radio Telescopes

The ONS25 telescope is used mainly for VLBI observations of, e.g., star forming regions, radio stars, and active galactic nuclei within the EVN. The ONS20 telescope is used for single-dish observations of astronomical objects and for astronomical (EVN and GMVA) and geodetic IVS experiments.  One recent major science VLBI highlight is the EVN localization of FRB121102 within about 10 mas (Marcote et al. 2017).

Examples of recent single-dish results using the ONS20 telescope include Wirström et al. (2016) who presented HCN observations of comets C/2013 R1 (Lovejoy) and C/2014 Q2 (Lovejoy), and Alatalo et al. (2016) who reported emission from multiple molecular species in NGC5195, a small disturbed companion galaxy to the large spiral M51, also using data from the CARMA interferometer.

### 3.11.    Mopra Radio Telescope

Mopra has been an essential part of the LBA network and has contributed to almost all VLBI publications from the southern hemisphere. It played a key role in the VSOP 5 GHz Active Galactic Nucleus Survey (Dodson et al. 2008). From the many years of mm-observing with the MOPS system, significant results have been obtained by a number of projects: the Millimetre Astronomy Legacy Team 90 GHz survey (MALT-90), aimed at characterising the physical and chemical evolution of high-mass clumps (Foster et al. 2011; Jackson et al. 2013; Rathborne et al. 2016); the $H_2O$ Southern Galactic Plane Survey (HOPS) at 22 GHz (Walsh et al. 2011); the spectral imaging survey of the Central Molecular Zone around the Galactic centre in 20 molecular lines from 85.3 to 93.3 GHz (Jones et al. 2012); and the Molecular Cloud Population of the Large Magellanic Cloud (MAGMA) LMC Survey (Wong et al.  2011).

### 3.12.    Korean VLBI Network

As a dedicated mm-VLBI facility, the scientific targets of the KVN include the inner regions of AGN jets, $H_2O$ and SiO masers from star formation regions and evolved stars. A legacy survey program related to the KVN calibrator survey (Lee et al. 2017) and four key science projects, two of them on AGNs and the other two on stellar masers, are also focused on those topics (https://radio.kasi.re.kr/kvn/ksp.php).

The multi-frequency simultaneous observing system of the KVN was developed to remove the non-dispersive tropospheric phase delay at mm wavelengths (Jung et al. 2011). This multi-frequency observing system opens the new possibility to detect faint sources at mm wavelengths with enhanced sensitivity, through use of the Frequency Phase Transfer (FPT) technique and the Source Frequency Phase Referencing (SFPR) technique (Rioja et al. 2015). The above-mentioned large projects of KVN benefit from FPT. Furthermore, SFPR provides accurate source positions between observed frequencies with enhanced detectability. This technique is used to explore opaque AGN cores (e.g. Jung et al. 2015) and to probe the circumstellar envelopes of evolved stars (e.g. Dodson et al. 2014).

KVN participates actively in international projects such as EVN, GMVA, and RadioAstron (e.g. Giovannini et al. 2018). Especially with VERA, KVN builds a joint array called KaVA and expands this collaboration to the East-Asian VLBI Network (An et al. 2018).

### 3.13.    VLBI Exploration of Radio Astrometry

Since VERA is dedicated to maser astrometry, nearly half of the VERA's machine time (i.e., 2000 hours per year) is spent on such observations, namely astrometric monitoring of target maser sources. The main targets are $H_2O$ masers in Galactic star-forming regions, and in some cases $H_2O$ and SiO masers in late-type stars are also observed. By measuring accurate positions of masers for a period of about one year, parallaxes and proper motions are obtained for target sources with an accuracy of the order of 10 μas. These data, combined with those obtained from other projects like VLBA/BeSSeL (Reid et al. 2014), are used to investigate the Galactic structure and to determine Galactic fundamental parameters, rotation curve, spiral arms and so on (e.g., Honma et al. 2012; Reid & Honma 2014; Reid et al. 2014).

In addition to studies of the Galactic structure, VERA is also used for the observation of individual target sources including AGN, late-type stars and star-forming regions. Regular monitoring of gamma-ray-emitting AGNs are on-going to trace their variability (see Nagai et al. 2013). Together with KVN, VERA takes part in the joint KaVA. Currently KaVA operates roughly 1000 hours per year, including open-use time and Large Programs (LP). As of 2017, three KaVA LPs are in operation, focusing on star-formation, AGB stars, and AGNs. Some initial results from KaVA's LPs are presented in Matsumoto et al. (2014), Yun et al. (2016), and Hada et al. (2017). Moreover, VERA takes part in the EAVN (East-Asian VLBI Network), a joint VLBI array involving China, Korea, and Japan. EAVN test observations have been done aiming at starting regular operations in late 2018.

## 4.    Characteristics of front-end receivers

In this section, an analysis is carried out for the technical properties of the receivers in operation at the various telescopes. For the fifteen radio telescopes included in the survey, 106 front-ends receivers operating in the frequency range 300 MHz - 116 GHz have been taken into consideration. GBT and EFF together operate 35% of these receivers (16 and 21, respectively). All the other radio telescopes have a number of receivers ranging between 4 and 10, apart for IRAM30, which is equipped with only one receiver at a frequency below 116 GHz.

Front-ends have been classified into the following categories: mono-feed; dual-feed; multi-feed; dual-frequency; UWB; and bolometer. Traditional mono-feed, dual-feed, and multi-feed differ in the number of parallel receiver chains, respectively one, two, and more than two. The dual-frequency category includes coaxial receivers which simultaneously observe two different frequency bands, while UWB indicates receivers with a frequency coverage much wider than traditional. Incoherent front-ends detecting only the incoming power are classified in the bolometer category. In this analysis, multi-frequency, non-coaxial receivers (like the KQW simultaneous receiver developed for KASI and now under consideration from other radio telescopes) are included in the mono-feed category.

Figure 2 illustrates the number of the receivers in operation at each radio telescope with the front-end category indicated in different colors. Mono-feed receivers are the most common, summing up to 70% of the total. Twelve receivers (11%) are dual-feed systems: four at GBT and EFF; two at TMRT; and one at MED and NOB. Multi-feed front-ends are a minority (7%), counting seven systems only: two at GBT and EFF; and one at SRT, NOB, and PARKES. The availability of dual- or multi-feed receivers is crucial for the execution of successful single-dish observations at high frequency, by allowing the efficient correction of atmospheric variations. Eleven receivers (10%) are simultaneous dual-frequency systems, the majority (7) of which are designed to cover the S and X bands for geodetic studies. SRT and TMRT operate dual-frequency receivers in the X and Ka bands for space science applications. Furthermore, two similar dual-frequency receivers for pulsar observations are installed at SRT and PARKES. Despite the large number of available front-ends at GBT and EFF, no dual-frequency system is installed at these facilities. Finally, an UWB front-end observing in the range 0.6-3 GHz and a bolometer working at 100 GHz are installed at EFF and GBT, respectively.

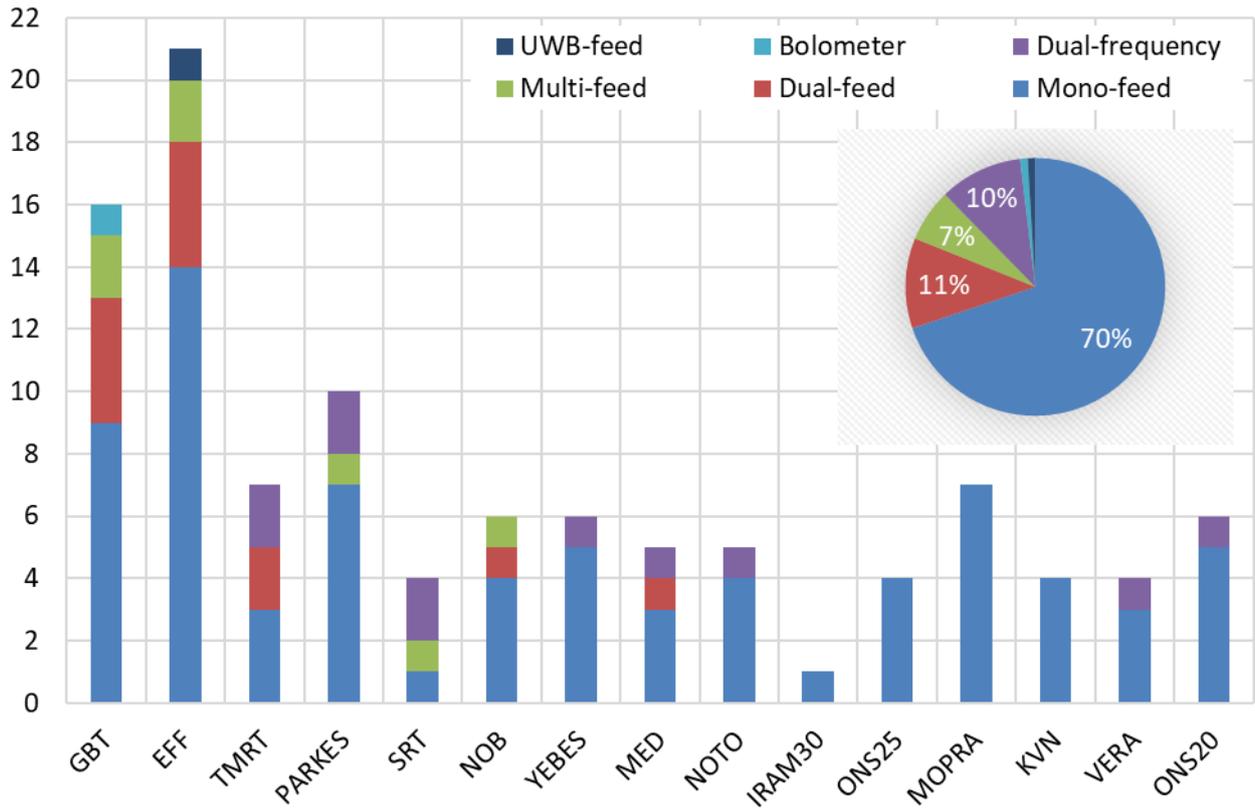

Figure 2 – Distribution of the receivers (divided by category) in operation. The panel on the right gives the cumulative distribution for each receiver category.

Fig. 3 summarizes the number of observed bands offered at each radio telescope. Based on its frequency coverage, each receiver has been included in one of the following frequency classes: below 1 GHz; 1-18 GHz; and above 18 GHz. For a given radio telescope, the total number of bands may be different from the total number of receivers either because the frequency range of a receiver crosses two classes or more than one receiver is available for the same frequency band. This last case happens in particular for GBT, EFF, and PARKES, due to the fact that either state-of-the-art receivers have been developed partially replacing the previous ones or each of them has been developed for a specific science case.

Despite the 1-18 GHz frequency range being the most widely available at all the radio telescopes, Fig. 3 points out a heterogeneous landscape, in terms of frequency classes, at the various radio telescopes. This may be attributed to several factors ranging from the technical characteristics of the radio telescope to the scientific interests of the radio astronomical community, motivating the construction of front-ends in some preferred frequency bands. Additionally, other non-negligible aspects are the site characteristics in terms of e.g. opacity and the local RFI environment. We notice that only five telescopes observe at frequencies below 1 GHz, four of them being large-size facilities (GBT, EFF, PARKES, and SRT), while only one belongs to the medium-size class (ONS25). As far as the higher frequency bands are concerned, we see that almost all the radio telescopes observe both in the 1-18 GHz and >18 GHz ranges. Exceptions are NOB and IRAM30, which are mm-facilities, and ONS25, which has no receiver working above 6.7 GHz.

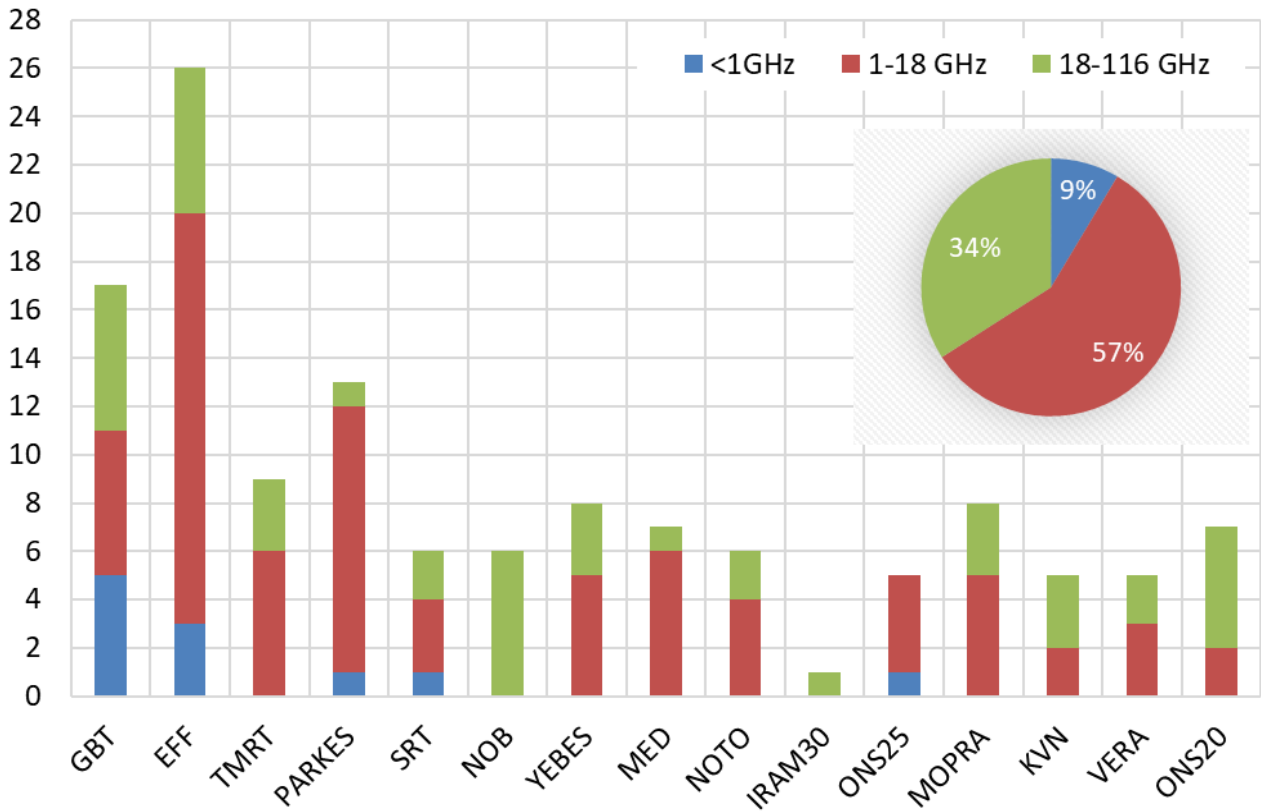

Figure 3 – Number of receiver bands in operation in the frequency range: < 1 GHz; between 1 and 18 GHz; and > 18 GHz. The panel on the right side gives the cumulative distribution for each frequency class.

The frequency distribution of the receivers in operation at different radio telescopes is given in Fig. 4 together with the indication of the categories defined at the beginning of this section. Remarkably, GBT and EFF are the only facilities offering an almost continuous frequency coverage between 300 MHz and 115 GHz.

It can be noted that there is a considerable amount of frequency overlap among the telescopes, in particular in specific bands like those traditionally used for single-dish spectral line studies or for VLBI observations. In fact, almost all the radio telescopes indicate the participation in the same international networks and projects, for instance the EVN and global VLBI, IVS, GMVA, EPTA, LEAP, and RadioAstron experiments. The majority of the facilities offer the L-, S/X-, C-, and K-bands and a significant number of radio telescopes can also observe in the Q- and W-bands. The existence of a common set of receivers, working in the same frequency bands at the various observatories could in principle allow the exchange of scientific projects among different radio telescopes (or sets of radio telescopes) provided a proper evaluation of the available sensitivity, angular resolutions, and observing time is done. Also, this could encourage the on-purpose organization of small antenna networks for VLBI-like observations of specific scientific projects.

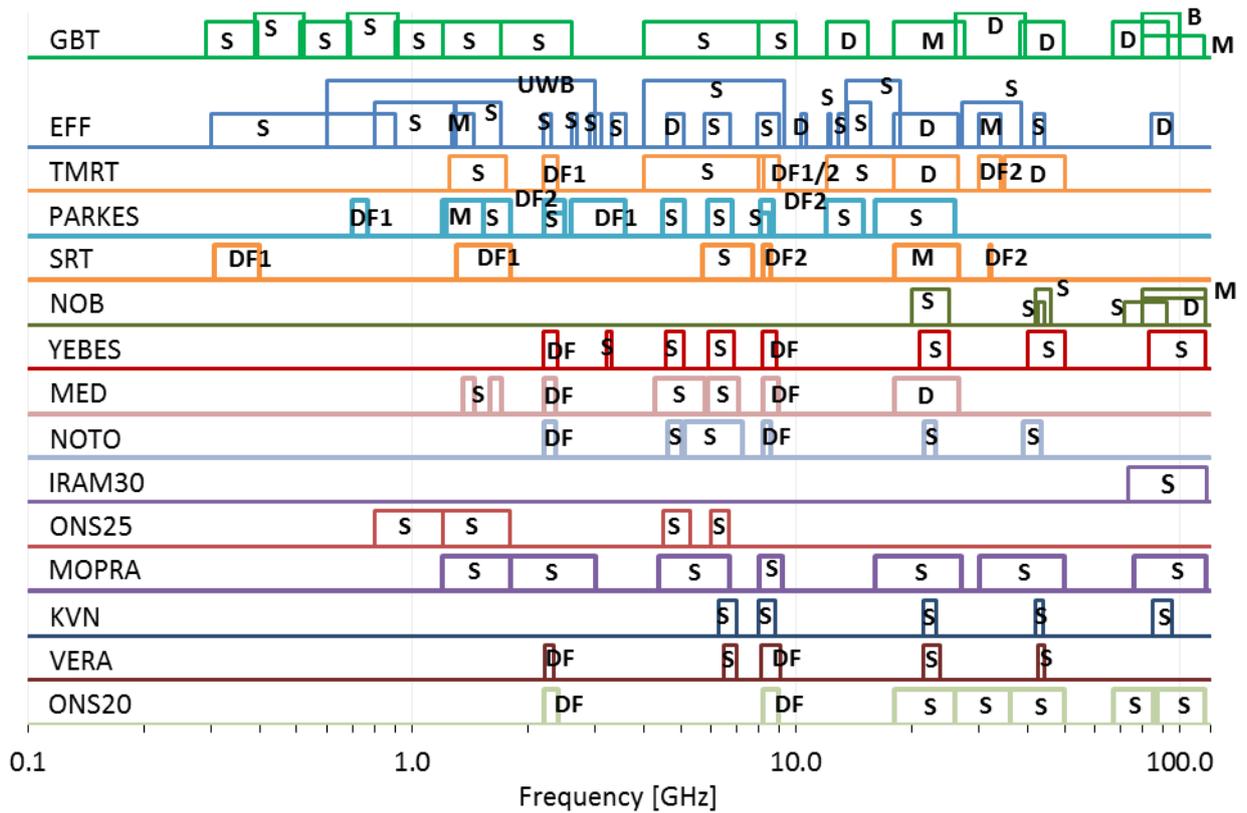

Figure 4 – Frequency coverage and category for operational receivers. Receiver categories are: S = mono-feed; D = dual-feed; M = multi-feed; DF = dual-frequency; B = Bolometer; and UWB = UWB-feed. For dual frequency receivers, the two bands are identified with the same number.

Figure 5 reports the year of construction for the receivers of the 13 telescopes which provided such information. This gives an indication of both the rate of front-end development for each facility and the average age of the receiver fleet for each site. For each telescope, the beginning of antenna operations is also indicated. The X/Ka-, K-, and S/X-bands receivers for SRT and NOTO are older than the telescope inauguration date because they were previously installed at MED. Almost all the GBT, TMRT, and YEBES receivers were available with full frequency coverage from the beginning of antenna operations, demonstrating a very efficient development and execution plan. We notice that three radio telescopes (MED, NOTO and EFF) are currently equipped with receivers older than 20 years. On the other side, recently NOB has fully renewed its receiver fleet.

As already reported, at the time of the survey (2017) EFF was the only telescope, among those taken into consideration, equipped with a UWB receiver. This front-end operates between 0.6 and 3.0 GHz and has been available since 2011, pioneering well ahead of time, with respect to other radio telescopes, this receiver category. The use of bolometer receivers started very recently (2017) with the W-band receiver in operation at the GBT. The first multi-feed receivers put in operation are those currently installed at EFF and SRT, dating back to the period 2005 – 2008. Basically, before 2005 all receivers were single-feed, dual-feed, or dual-frequency architectures.

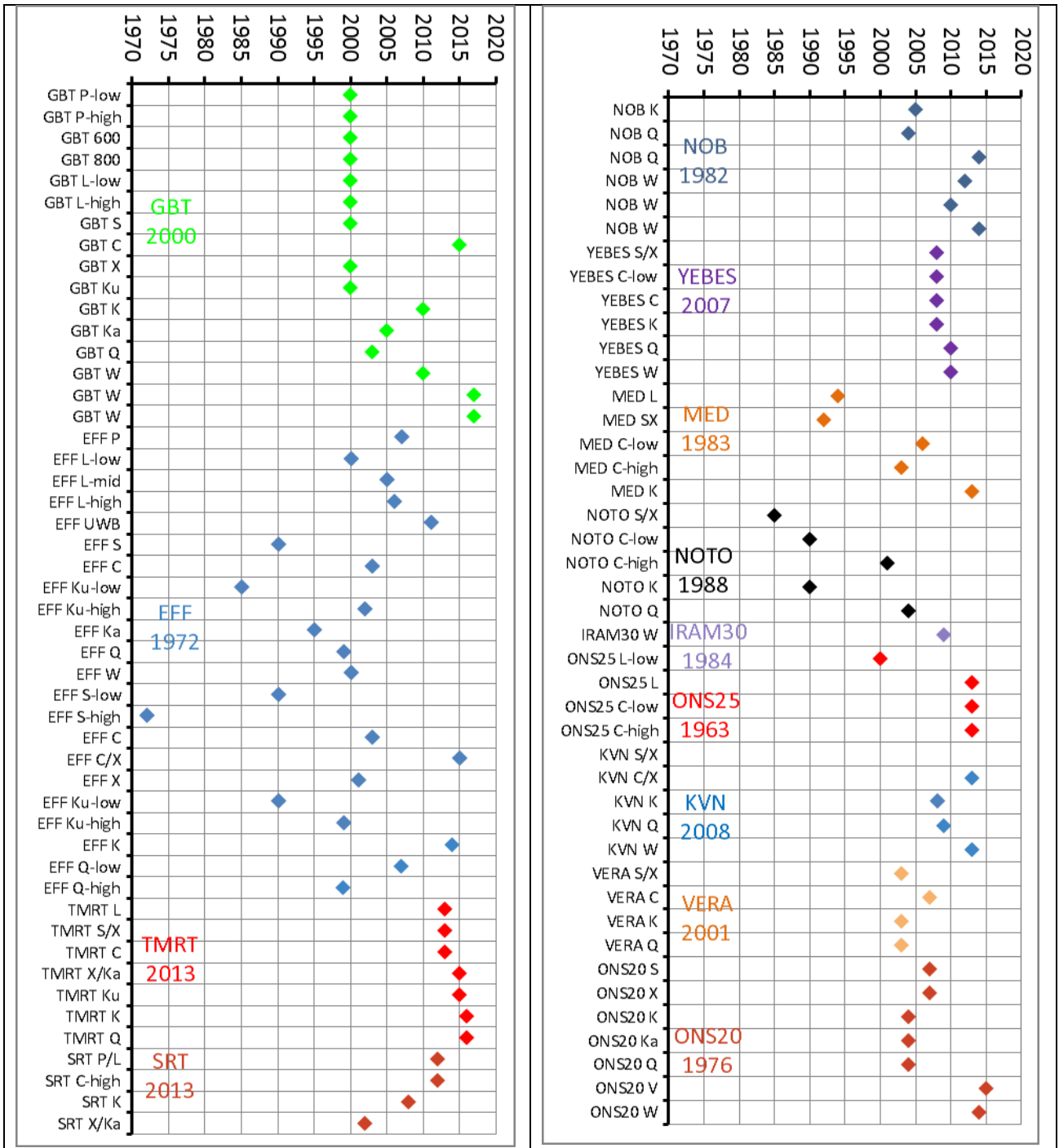

Figure 5 – Age of operational receivers. The years close to the radio telescope abbreviation indicates its inauguration date.

One of the more significant key parameters in the front-end performance is the receiver noise temperature. Figure 6 reports for most of the front-end receivers included in this survey the receiver noise temperature versus the central frequency of the front-end. Different colours are associated with the various radio telescopes. For each frequency band, a large scatter (up to one order of magnitude) in the front-end receiver noise temperature is visible and is due to the intrinsic technical characteristics implemented in the receivers, such as the LNA's noise temperature, the physical temperature of the receiver, and the losses in the first components of the feed-system. The actual performance of the receivers on the telescope also depends on a number of other factors, as described in the next section.

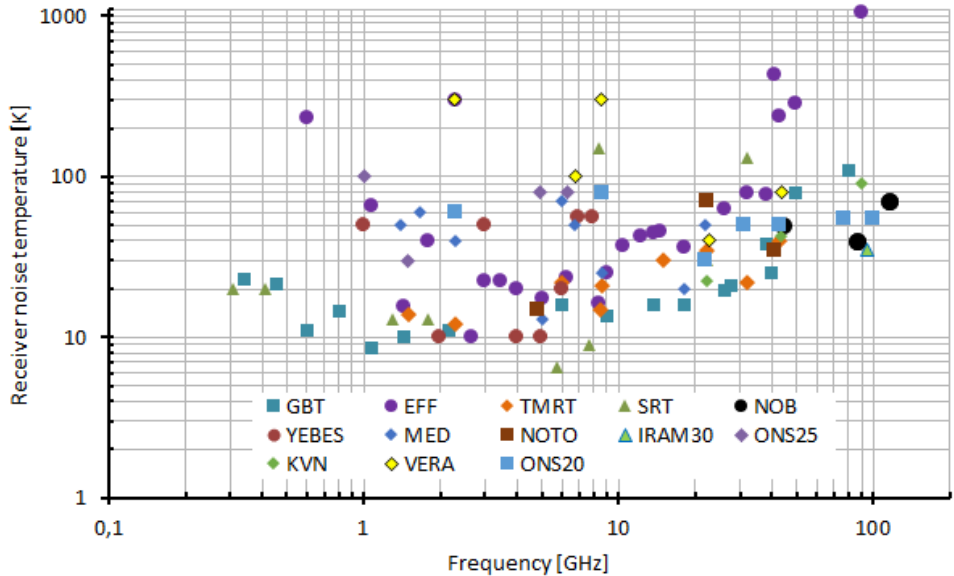

Figure 6 – Receiver noise temperature for the operational receivers.

## 5.  Observing performance of Radio Telescopes

In this section, we discuss the performance of the fifteen radio telescopes involved in the survey by using the receiver characteristics described in section 4. A convenient parameter identified for this comparison is the System Equivalent Flux Density (SEFD), which is expressed as the ratio between the overall system noise temperature (including receiver noise temperature and the atmospheric contribution) in Kelvin and the antenna gain in Kelvin/Jansky. In this analysis the receiver performance at the central frequency is taken as representative of the whole bandwidth. However, when the performance varies considerably inside the band, the values at the band edges are plotted if available.

It must be noted that some factors not related to the intrinsic properties of the front-ends affect the comparison of their performance. The relative gain of two antennas with different collecting area scales as the squared ratio of the antenna diameters, therefore a very well-designed front-end with excellent noise performance may have a moderate SEFD because of the limited collecting area of the radio telescope on which it is mounted. Also, the overall surface accuracy of the radio telescope may degrade the antenna gain. For example, the combination of these two factors is such that the peak gain of the 22 GHz multi-feed receiver is 0.66 K/Jy if mounted on SRT, while at MED it decreases to 0.11 K/Jy.

To allow for a fair comparison among telescopes with similar collecting area, in Figs. 7 and 8 the SEFD of telescopes in the large-size and medium-size classes respectively are shown. It can be seen most of SEFD values for large-class radio telescope vary between 10 and 300 Jy, while for the latter class from 100 to 3000 Jy.

The most evident result of Fig. 7 is that GBT has the best performance, with SEFD values distributed within only one order of magnitude over the entire frequency coverage of the available receivers. The large collecting area of EFF permits SEFD values comparable to the GBT in the 2-10 GHz frequency range, while the worse surface accuracy and the older receiver suite account for a deterioration of the performance at respectively the higher and the lower frequencies. Two other large-class telescopes, TMRT and SRT, which are similar in terms of construction date and technical features, show indeed similar SEFD properties. They both have lower performance than EFF due to their smaller size. The last large-class radio telescope, PARKES, despite its long career as an observational facility, remarkably shows performance comparable to the more modern TMRT and SRT up to K band.

Among the middle-size class telescopes (Fig. 8), YEBES shows the best performance up to the K band, while at higher frequencies NOB has the best SEFD. NOB site altitude and high surface accuracy reflects in SEFD values at 100 GHz comparable to those of the large-size class EFF radio telescope. The antennas of the VERA array have the highest SEFD due to the relatively high noise receiver temperature (see Fig. 6). The antennas of the other interferometers considered in the survey, KVN, show performance in line with other single-dishes like MOPRA and ONS20. The other medium-size class radio telescopes have similar SEFD performances.

Overall, for both telescope classes the receiver SEFDs show an almost flat trend up to the K-band and increase at higher frequencies as expected, due to the contribution of the atmosphere.

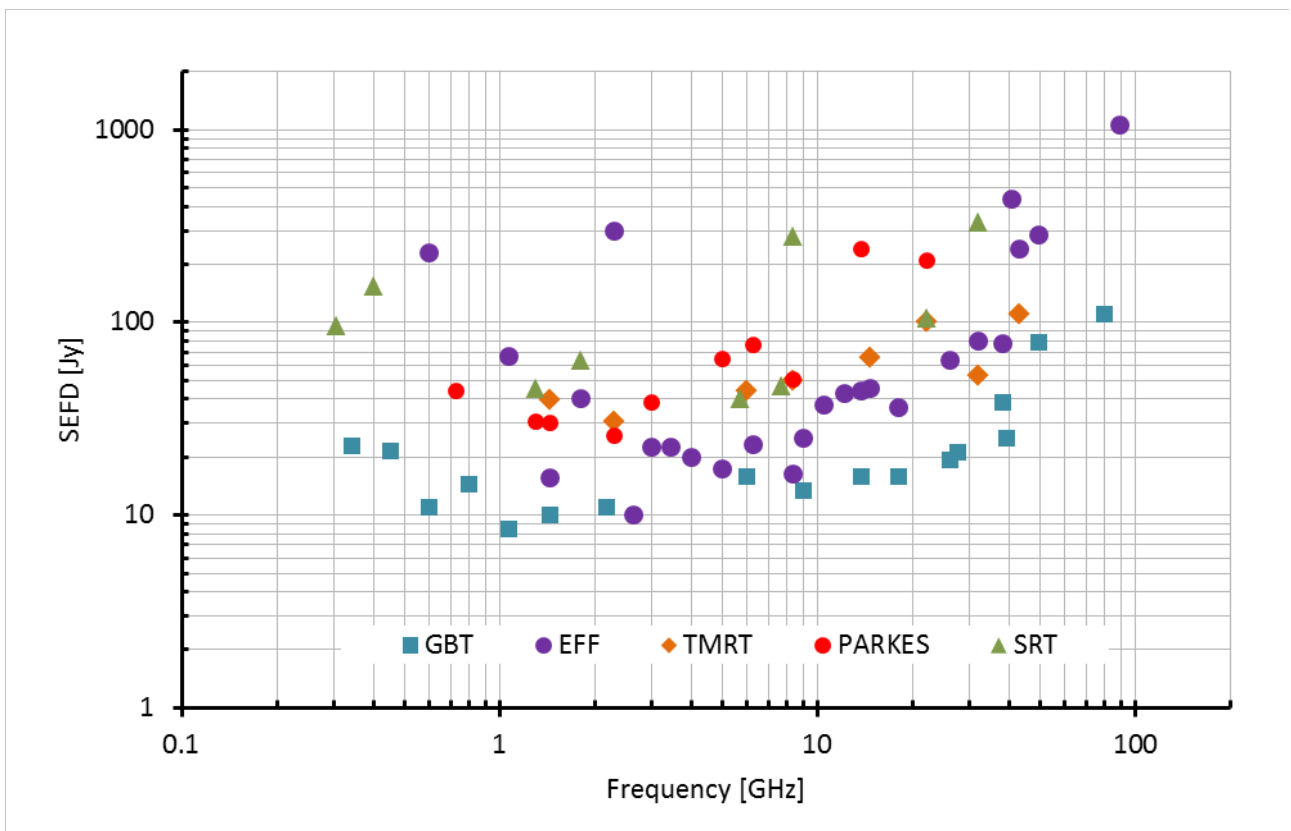

Figure 7 – SEFD for operational receivers at large-size class radio telescopes.

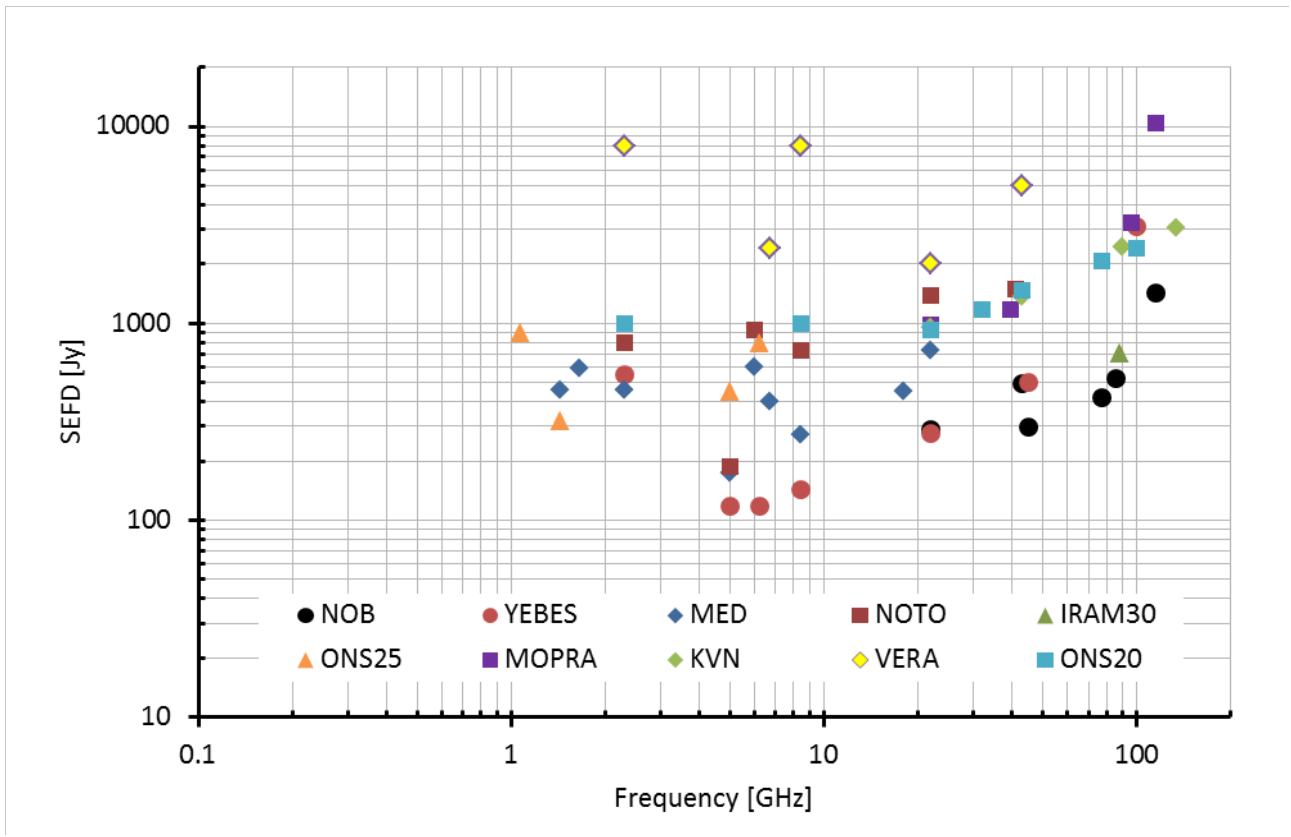

Figure 8 – SEFD for operational receivers at medium-size class radio telescopes.

The antenna size is the dominant factor when comparing SEFD values. To highlight the contribution of other terms, we applied a normalization factor related to the telescope diameter. In Figs. 9 and 10, the distribution of the SEFD normalized with respect to a reference antenna of diameter 32 m is shown. In other words, each SEFD value has been multiplied by the ratio:

$$\left(\frac{diameter}{32}\right)^2$$

where *diameter* is the diameter (in meters) of the telescope hosting that receiver.

The majority of normalized SEFD values for medium-size class antennas are in the range 200-3000 Jy while values for the large antennas range from 100 to 3000 Jy. These very similar ranges indicate that the main reason for the varying performance seen in Figs. 7 and 8 is due to the telescope collecting area. The remaining differences in the normalized SEFD are due to other characteristics like surface accuracy, offset antenna type, and system noise temperature. Even in the normalized plot, the GBT is the telescope characterized by the best performance demonstrating a successful handling of the above characteristics in the instrument development phase. In particular, we point out the GBT unblocked aperture, active surface and low noise temperature of the receiver suite. Furthermore, from the normalized SEFD analysis we note that even low altitude sites, like ONS20 and KVN, demonstrate good performance at frequencies where atmospheric contributions play a significant role.

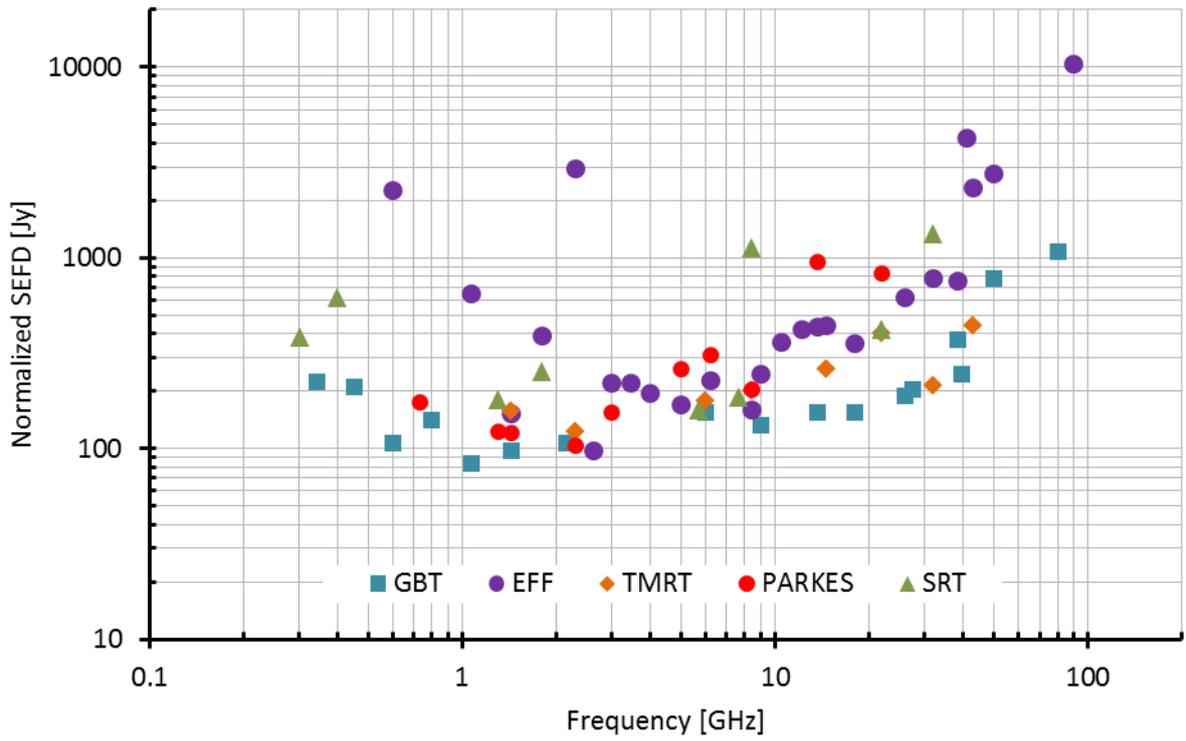

Figure 9 – Normalized SEFD for operational receivers at large-size class radio telescopes.

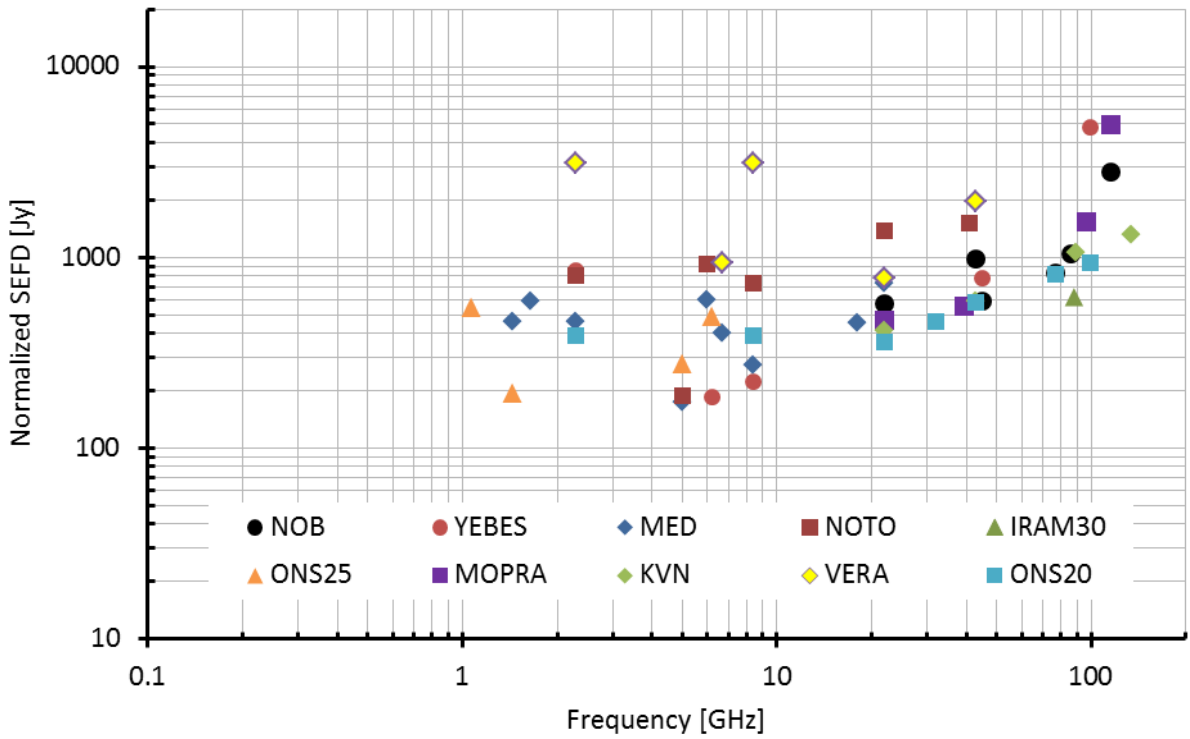

Figure 10 – Normalized SEFD for operational receivers at medium-size class radio telescopes.

A fundamental parameter describing the scientific performance of a radio telescope is the sensitivity, defined here as the 1-sigma RMS noise in mJy that can be reached within the nominal instantaneous bandwidth in 1 second of integration time. The sensitivity thus takes into account also the back-end characteristics in terms of instantaneous bandwidth that can be processed. Figures 11 and 12 show the sensitivity for large-size and medium-size telescopes, respectively. Sensitivity calculations have been made using the bandwidth of the total-power back-end as specified in the technical information provided by each observatory.

Remarkably, the overall shape of the distribution of sensitivity values for both radio telescope classes is very similar to that of the SEFD, indicating similar capabilities in terms of the available instrumentation at almost all the observatories.

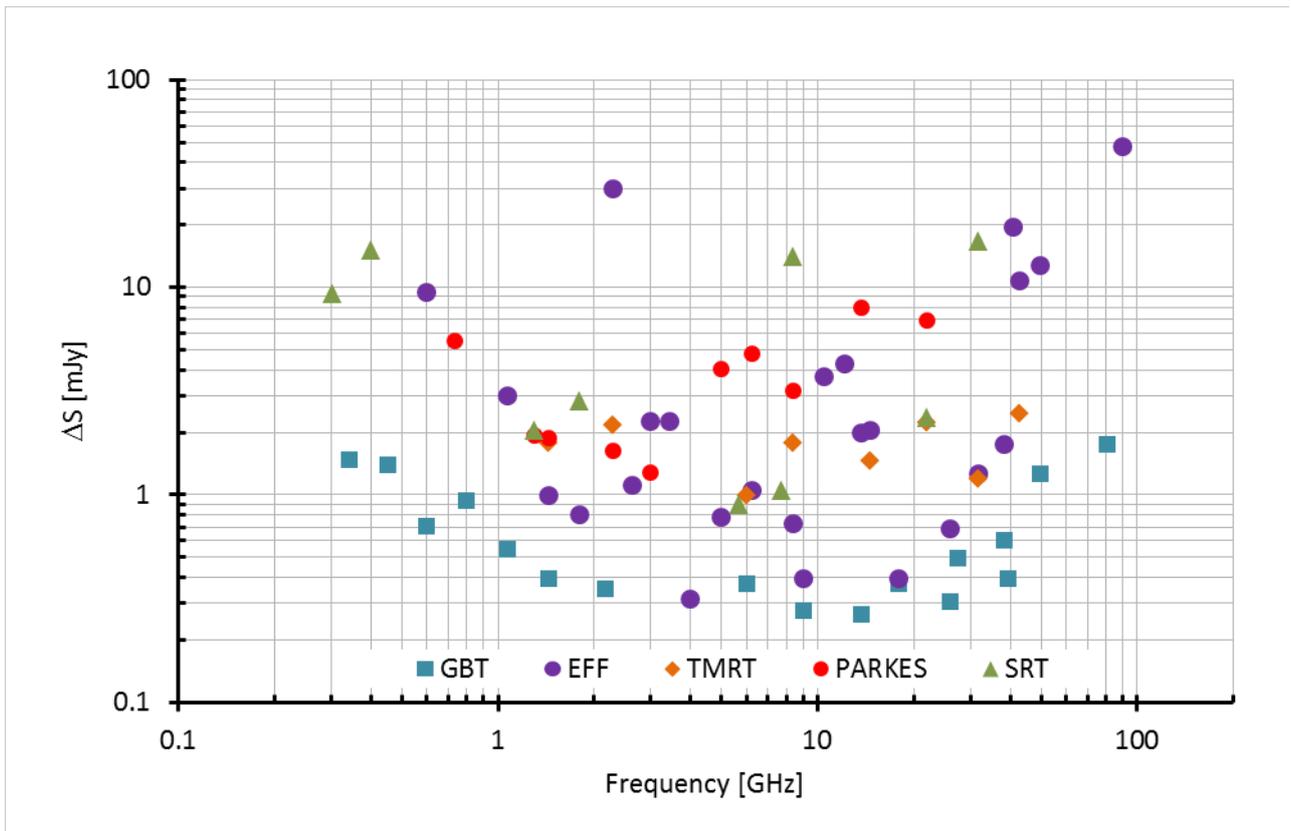

Figure 11 – Sensitivity for operational receivers at large-size class radio telescopes.

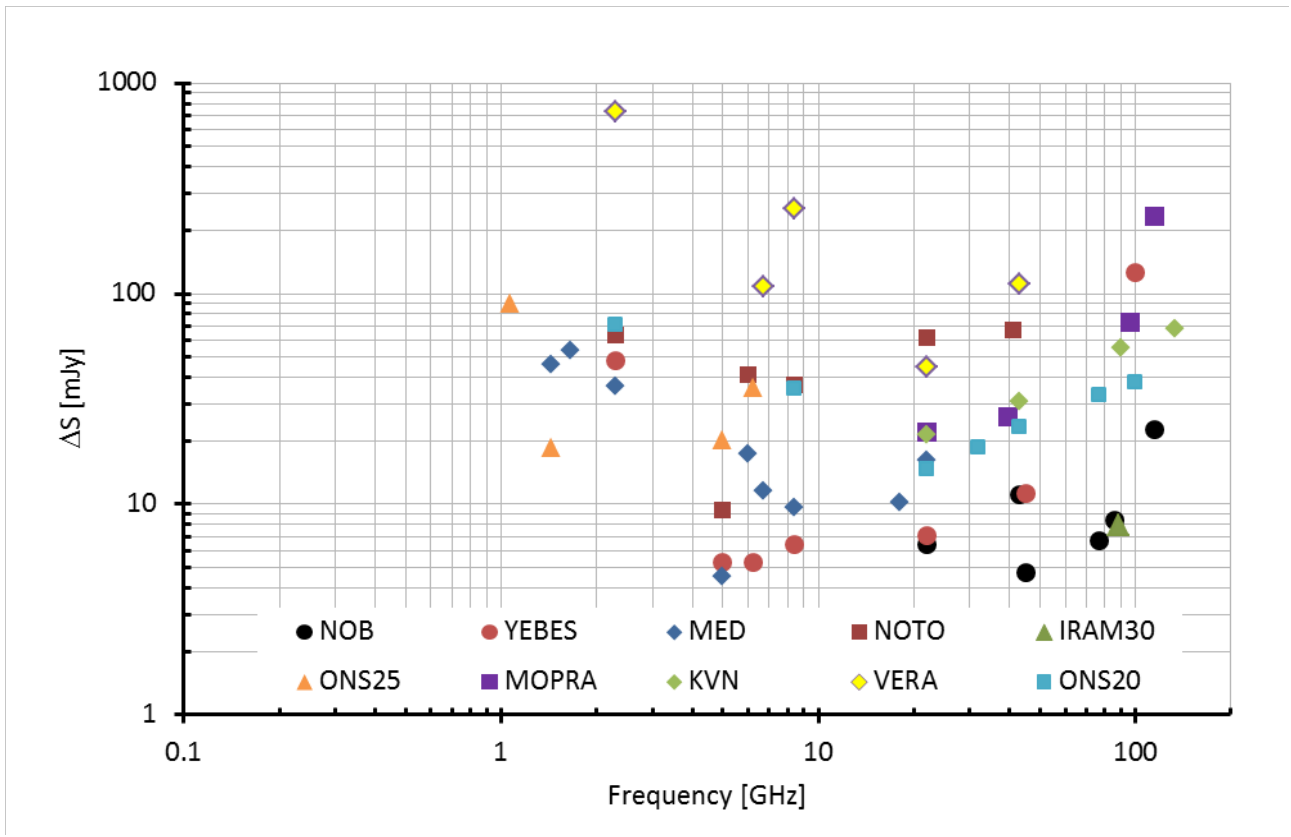

Figure 12 – Sensitivity for operational receivers at medium-size class radio telescopes.

## 6.    Future trends in front-end receivers development

From the analysis of the 22 receivers currently under development at the radio observatories included in this analysis, a significant change in the front-end architecture is registered. The characteristics of these receivers in terms of frequency coverage and feed category are summarized in Fig. 13. Out of 22 receivers, traditional front-ends count for eight of mono-feed architecture, four dual-feed, three multi-feed, and one dual-frequency. The exploitation of new technologies appear in the development plan for two low frequency PAF systems and five UWB solutions. From this perspective, it is interesting to note that PARKES is going towards a completely renewed receiver suite composed only by UWB and PAF technologies. In terms of frequency distribution, the 1-18 GHz frequency band is the one with the largest number of receivers under development. Some of the large-size radio telescopes are constructing new front-ends at frequencies below 1 GHz, while the SRT is planning to install a 100-GHz receiver. At the time of the review, YEBES is the only medium-class telescope developing receivers for simultaneous high-frequency observations.

The GBT is strongly pushing towards development in the W-band with different front-end typologies: a 50x multi-feed and a PAF solution. Both these projects are under discussion and therefore not illustrated in Fig. 13.

Dual-feed receivers are the most common at high frequencies, despite the availability of technological skills for developing multi-feed devices. This might be due to a number of reasons ranging from economical considerations with respect to the scientific drivers and technological challenges posed by such development.

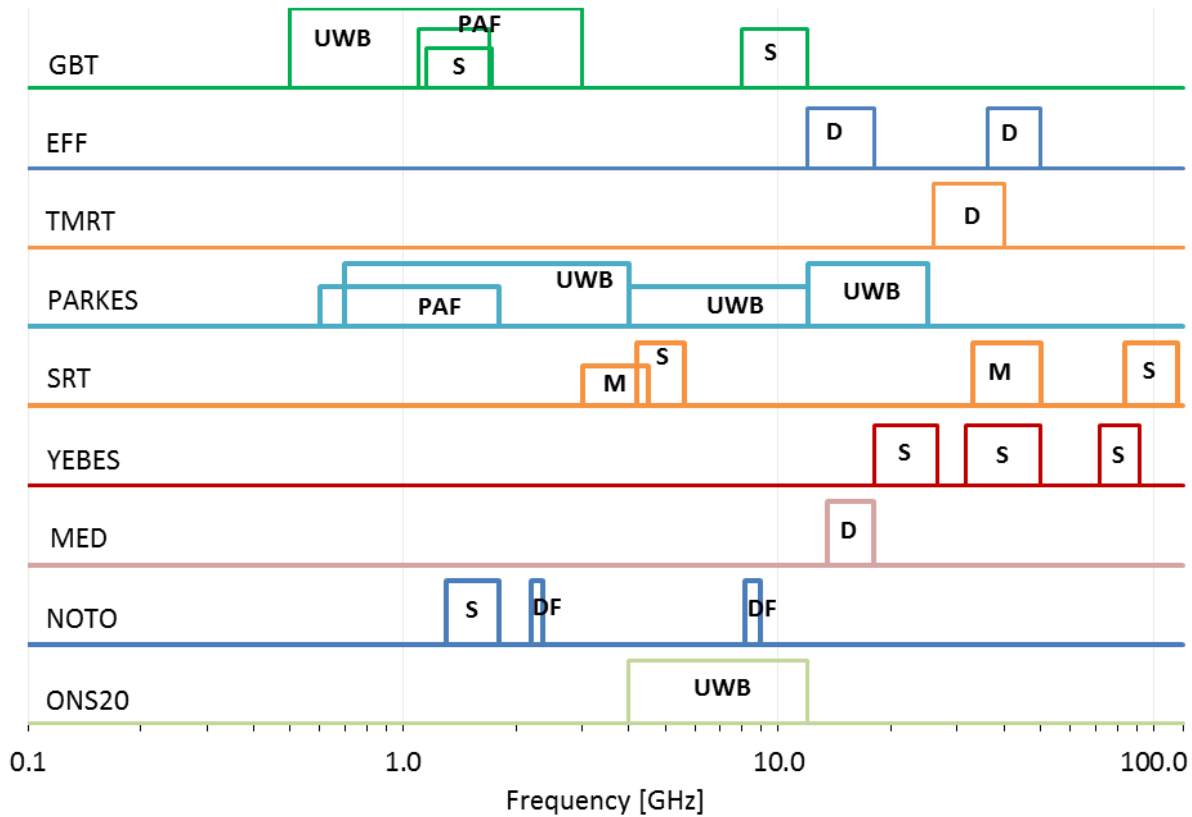

Figure 13 – Frequency coverage and category for receivers under development. S = mono-feed; D = dual-feed; M = multi-feed; DF = dual-frequency; PAF = phased array feed; and UWB = UWB-feed.

## 7.    Conclusions

In the previous sections, we have seen that the receivers currently in operation at a selection of radio telescopes are mainly composed (90% of the total) by single-feed, dual-feed, and dual-frequency solutions. In the early 2000s, multi-feed receivers started to be constructed thanks to the ability to serialize the fabrication of microwave components and to pack them into the dewar. Nowadays, this category covers 7% of the total receivers in operation. Apart from a couple of exceptions, no other types of receiver are currently operated on the radio telescopes selected in this survey.

With the exception of the GBT, EFF, and PARKES radio telescopes, all the other facilities are equipped on average with 6 to 8 receivers. Most of them include the frequencies typical of VLBI, but with larger bandwidths with respect to that strictly required for VLBI experiments in order to allow also efficient single-dish observations. This points out that modern receivers are usually built to be general purpose instruments, to make it possible to address different science cases (in principle also beyond those considered when defining the receiver specifications).

The availability of many front-ends puts strong constraints on the efficiency of the radio telescope in changing receivers. Therefore, most of the antennas are currently equipped with a frequency agility system. This system is a key element to allow dynamic scheduling at the observing facilities i.e. the capability to recognize the best observing frequency depending on the current atmospheric opacity. Frequency agility can thus be seen as a requisite that will be more and more fundamental for future instrumental developments. In the last decades, remarkable progress has been made in the construction of receiver hardware components, for instance Low Noise Amplifiers. This has led to significant improvements in the receiver noise temperature and available bandwidth, and allowed for the first time the construction of UWB receivers. UWB receivers are able to simultaneously observe many frequency bands, in different observing modes (continuum and spectroscopy) and have also the advantage of limiting the maintenance and the operational costs of the related infrastructure. PARKES, for example, is expected to fully replace the traditional single-feed receiver fleet with new UWB systems (for a comparison see Figs. 4 and 13). Also, the RadioNet H2020 project is developing the UWB BRAND prototype to be installed on board one of the European radio astronomical facilities. The adoption of UWB receivers put strong requirements on back-end capabilities and on RFI mitigation strategies. The next challenge will be to combine UWB receivers and multi-feed technologies to realize receivers capable of simultaneously and efficiently observing large bandwidth and extended areas of the sky.

Another technological frontier is represented by the development of PAF receivers, currently available at some radio telescopes for low frequency observations and for which many research groups worldwide are investing significant resources. This new technology looks very promising and its time-line appears aligned with Phase 2 of the SKA, which may mount such receiving systems on its dish elements.

In parallel with these new areas of technological development, this survey has shown that the era of traditional receivers is not concluded in terms of benefiting from technological advances in the microwave active and passive devices.

As a final consideration, we point out that the main technological breakthroughs in the area of front-end developments are realized within joint international collaborations, where each radio observatory brings its individual experience and shares it with other partners. Such a working model goes in the same direction as that adopted by the next generation radio telescopes, like the SKA, which involves the worldwide astronomical community in the realization of cutting-edge scientific projects.

## Acknowledgements

This work is a follow-up of the front-end review process for the Italian radio telescopes sponsored by the INAF Scientific Directorate.

The Green Bank Observatory is a facility of the National Science Foundation operated under cooperative agreement by Associated Universities, Inc.

The Effelsberg 100-m radio telescope is operated by the Max-Planck-Institute for Radio Astronomy in Bonn on behalf of the Max-Planck-Society.

The TianMa Radio Telescope is operated by Shanghai Astronomical Observatory of Chinese Academy of Sciences.

Since 2013, the IRAM30 radio telescope is part of the Spanish ICTS (Singular Scientific and Technical Infrastructures) network.

The Parkes and Mopra radio telescopes are part of the Australia Telescope National Facility, which is funded by the Australian Government for operation as a National Facility managed by CSIRO.